\documentclass{aa}
\usepackage{graphicx}
\usepackage{txfonts}
\newcommand\vv{{\mathrm v}  }       

\begin{document}
   \title{Eccentric binaries}

    \subtitle{Tidal flows and periastron events}

   \author{E. Moreno  
       \inst{1}
   G. Koenigsberger 
       \inst{2}
   \and
   D. M. Harrington
       \inst{3}
        }

   \offprints{G. Koenigsberger}

\institute{Instituto de Astronom\'{\i}a, Univer\-sidad Nacio\-nal
           Aut\'o\-noma de M\'exico, M\'exico D.~F. 04510, Mexico\\
           \email{edmundo@astroscu.unam.mx}
\and
           Instituto de Ciencias F\'{\i}sicas, Univer\-sidad Nacio\-nal Aut\'o\-noma de M\'exico, Cuernavaca, Morelos, 62210, Mexico\\
           \email{gloria@astro.unam.mx}
      \and
           Institute for Astronomy, University of Hawaii, {2680 Woodlawn Drive}, Honolulu, HI, 96822\\
           \email{dmh@ifa.hawaii.edu}
          } 
 \date{Received; accepted }

\abstract
{A number of binary systems present evidence of enhanced activity around periastron passage, 
suggesting a connection between tidal interactions and these periastron effects.}
{The aim of this investigation is to study the  time-dependent response of a star's surface 
as it is perturbed by a binary companion.  Here we focus on the tidal shear energy dissipation.}
{We derive a mathematical expression for computing the rate of dissipation, $\dot{E}$,
of the kinetic energy by the viscous flows that are driven by tidal interactions on the
surface layer of a binary star. The method is tested by comparing the results from
a grid of model calculations with the analytical predictions of Hut (1981) and the
synchronization timescales of Zahn (1977, 2008). 
 }  
{Our results for the dependence of the average (over orbital cycle)  energy dissipation,
$\dot{E}_{ave}$, on orbital separation are consistent with those of Hut (1981) for 
model  binaries with an  orbital separation at periastron $r_{per}/R_1 \gtrsim 8$,
where $R_1$ is the stellar radius.
The model also  reproduces the predicted pseudo-synchronization angular velocity for 
moderate eccentricities ($e\leq$0.3).  In addition,  for circular orbits our approach yields the same
scaling of synchronization timescales with orbital separation as given by Zahn (1977, 2008)
for convective envelopes.  The computations  give  the distribution of $\dot{E}$ 
over the stellar surface, and show that  it is generally concentrated
at the equatorial latitude,  with maxima generally located around four clearly defined 
longitudes,  corresponding to the fastest azimuthal velocity  perturbations.
Maximum amplitudes occur around periastron passage or slightly thereafter for supersynchronously rotating
stars. In very eccentric binaries,  the distribution of $\dot{E}$ over the surface
changes significantly as a function of orbital phase, with small spatial structures appearing 
after periastron.  An exploratory calculation for a highly eccentric binary system with
parameters similar to those of  $\delta$ Sco ($e$=0.94, $P$=3944.7 d) indicates that $\dot{E}_{ave}$ 
changes by $\sim$5 orders of magnitude over the 82 days  before  periastron, suggesting that
the sudden and large amplitude variations in surface properties around periastron may, indeed, contribute
toward the activity observed  around this orbital phase.
}
{}
  \authorrunning{Moreno, Koenigsberger \& Harrington}
  \titlerunning{Tidal flows and periastron events}
   \maketitle

\section{Introduction}\label{introd}

A number of binary systems present evidence of enhanced activity around
periastron passage. Among these, $\eta$ Car and the Wolf-Rayet systems
WR 140 and WR 125 are the most extreme and best documented examples
of periodic brightening at X-ray, visual and IR wavebands associated
with periastron passages. Recently, van Genderen \& Sterken (2007)
 suggested that these periastron events may have the same
physical cause as the milder ``periastron effects" exhibited by many
renowned eccentric binaries in which small enhancement
($\Delta$m$_v\sim$0.01--0.03$^{mag}$) in the visual brightness of the
system around periastron passage are observed. They suggest that the
fundamental cause of the effects may reside in the enhanced tidal
force that is  present during periastron passage. In addition to brightness enhancements,
binary interactions are frequently invoked to explain certain mass-ejection phenomena.
For example,  Koenigsberger, Moreno \& Cervantes (2002)
raised the question of whether tidal forces play a role in changing the wind structure
in the massive Small Magellanic Cloud binary system HD 5980, and suggested a possible 
link between tidal forces  and the instability producing the 1994 eruptive event. 
 Bona\`ci\'c Marinovi\'c et al. (2008) proposed a tidally-enhanced model
of mass-loss from AGBs. Observational data supporting the idea of  mass-loss events
associated with periastron passage exist for the highly eccentric ($e\sim$0.94)
binary $\delta$ Sco (Bedding 1993; Miroshnichenko et al. 2001; Tango et al. 2009).
And Millour et al. (2009) suggested that the dusty circumbinary environment in HD 87643
might be associated with repeated close encounters in the long period and possible
highly eccentric binary.   In this paper we explore a 
mechanism that has the potential of providing a theoretical framework for 
analyzing these phenomena.

Tidal interactions are ubiquitous in binary systems of all types.
They lead to the deformation of the stellar surface. When the system
is in equilibrium\footnote{A binary star is defined to be in
equilibrium (Hut, 1980) when it is in a circular orbit (e=0),
its rate of equatorial rotation equals the rate of orbital motion
($\omega_{rot}=\Omega$) and the axes of stellar and orbital
angular velocity are both perpendicular to the orbital plane.},
the deformation remains constant. However, if the stellar rotation
is not synchronized with the orbital motion, the shape of the star is time-dependent.
In eccentric binaries, the orbital separation changes as the stars move
from periastron to apastron, thus leading to a time-dependent gravitational 
force. In addition, because the orbital angular velocity, $\Omega$, is a 
function of the orbital separation,  the departure from synchronicity between
the stellar rotation angular velocity, $\omega$, and $\Omega$ is orbital-phase
dependent.   Hence, eccentric binaries are {\em never} in synchronous rotation.
The asynchronicity leads to the appearance of potentially large 
horizontal motions on the stellar surface, which are referred to as 
{\it tidal flows}.  

The energy that is dissipated because of shearing motions is ultimately
converted into heat that is deposited in the stellar layers.  Whether
this process leads to detectable observational effects is one of the 
questions that has driven our investigation of the detailed tidal interaction effects.
In the first paper of this series,  a very simple model was presented
for computing the stellar surface motions in a component of a
binary system (Moreno \& Koenigsberger 1999, 
hereafter Paper I).  In that model we computed only the motion of surface 
elements on the equatorial plane of a star.  We followed a Lagrangian
approach in a quasi-hydrodynamic scheme, solving the equations of
motion of small surface elements, as they respond to gravitational,        
gas pressure, viscous, centrifugal, and Coriolis forces. The model         
was applied to the ${\iota}$ $Ori$ binary system (Paper I), the $\epsilon$ 
$Per$ system (Moreno, Koenigsberger \& Toledano 2005, hereafter Paper II),
and the optical counterpart of the X-ray binary 2S0114+650 (Koenigsberger et al. 
 2006). In Paper II we computed the absorption line profiles obtained with the model,
taking an ad-hoc extension to stellar polar angles of the motion
computed on the stellar equator. This extension was also used in the
analysis of energy dissipation rates and synchronization time scales
in binary systems with circular orbits (Toledano et al. 2007). Here 
we improve the model by computing explicitly the motion of elements along
different parallels that cover the stellar surface. 
This new scheme in the computation of energy dissipation was applied in the LBV/WR 
system HD 5980 (Koenigsberger \& Moreno 2008), and in the study of tidal 
flows in $\alpha$ Virginis (Harrington et al. 2009).

This paper is organized as follows:
In Section 2 we generalize the mathematical expression for the calculation 
of shear energy dissipation rates, $\dot{E}$, that was presented in  Toledano 
et al. (2007); in Sections 3 and 4 we compare the predictions of our model 
for $\dot{E}$ with the analytical results of Hut (1981) and the synchronization
timescales with the analytical expression of Zahn (1977, 2008), respectively;    
in Section 5 we explore the orbital-phase dependent effects and their possible
relation to the observational periastron effects;   in 
Section 6 we present a discussion, and Section 7 lists the conclusions.

\section{The extended model}\label{modeloext}  

As in the first version of the model, we assume that the main stellar
body, below the thin surface layer, behaves as a rigid body.
In the quasi-hydrodynamic method used in Paper I, for every surface 
element a detailed analysis of the positions of neighboring elements is
needed to assign a local dimension. If an initial well ordered
arrangement of the elements begins to mix freely in the azimuthal and
polar directions, the assignment of a physical dimension to an element
and its corresponding interactions with neighboring elements is not
treatable in our current scheme.  Thus, we make the simplifying assumption
that the surface elements move only in the radial and azimuthal
directions; that is, an element always stays in its initial parallel, and there is
no meridional motion.\footnote{Eqs. 40b and 40c given in Scharlemann (1981) 
indicate that for a colatitude angle $\theta>$70$^\circ$, the maximum perturbation
in the azimuthal direction is more than 15 times greater than the maximum 
perturbation in the polar direction.}
The  energy dissipation computation performed  in Section \ref{disip} requires 
the motions of the surface elements. In this section 
we give the procedure to compute this motion in either of the two stars in the  
binary system, say star 1, with mass $m_1$ and initial radius $R_1$. 
Star 2 has a mass $m_2$ and instantaneous position
{\boldmath $r$}$_{21}$ with respect to the center of $m_1$, and an orbital
angular velocity $\bf {\Omega}$. The motions of the surface elements
in $m_1$ are computed in a primed non-inertial reference frame with
its origin at the center of $m_1$, and Cartesian axes rotating with the
orbital angular velocity $\bf {\Omega}$; the $x'$ axis always points
to $m_2$. 

In Paper II the total acceleration {\boldmath $a$}$'$ of a
surface element, measured in the non-inertial frame, is shown to be

\begin{eqnarray} \mbox{\boldmath $a$}'& =& \mbox{\boldmath $a$}_{\star}-
 \frac{Gm_1 \mbox
{\boldmath $r$}'}{|\mbox{\boldmath $r$}'|^3}-Gm_2 \left [
\frac{\mbox{\boldmath $r$}'- \mbox{\boldmath $r$}_{21}}
{|\mbox{\boldmath $r$}'-
\mbox{\boldmath $r$}_{21}|^3}+ \frac{\mbox{\boldmath $r$}_{21}}
{|\mbox{\boldmath $r$}_{21}|^3} 
\right ]- \nonumber \\ 
&& - \mbox{\boldmath ${\Omega}$} \times \left ( \mbox{\boldmath
 ${\Omega}$} \times
\mbox{\boldmath $r$}' \right ) 
-2 \mbox{\boldmath ${\Omega}$} \times \mbox{\boldmath $\vv$}'- 
\frac{d \mbox{\boldmath ${\Omega}$}}{dt} \times \mbox{\boldmath $r$}',
\label{ecmov} \end{eqnarray}

\noindent with {\boldmath $a$}$_{\star}$ the acceleration of
a surface element produced by gas pressure and viscous forces exerted by 
the stellar material surrounding the element, and {\boldmath $r$}$'$,
{\boldmath $\vv$}$'$ the position and velocity of the element in the
non-inertial frame. A nearly spherical shape is assumed for $m_1$
throughout its motion. 

The computation starts at time $t = 0$  with no relative motion between the surface
elements, so that  at this initial time the acceleration
{\boldmath $a$}$_{\star}$, with value
{\boldmath $a$}$_{{\star}o}$, has only the contribution of gas
pressure. To compute this acceleration we consider a second, 
double-primed, non-inertial reference frame, whose origin is also at
the center of $m_1$, and has axes rotating with the assumed constant 
angular velocity {\boldmath ${\omega}$}$_{\star}$ of the inner rigid
region of star 1. This angular velocity is written as
{\boldmath ${\omega}$}$_{\star}$ = ${\beta}_0$$\bf {\Omega}$$_0$, that 
is, equal to a certain fraction ${\beta}_0$ of the initial orbital 
angular velocity.\footnote{The expression  
$\beta_{per}=\omega_0/\Omega_{per}=$ 0.02 P $\frac{\vv_{rot}(1-e)^{3/2}}{R_1(1+e)^{1/2}}$
is a convenient form to compute the synchronicity parameter at periastron.  
$P$ is in days, $\vv_{rot}$ in km s$^{-1}$, and $R_1$ in $R_\odot$.}  
The condition for initial equilibrium {\boldmath $\vv$}$''_0$ = 0, {\boldmath $a$}$''_0$ = 0 
on the surface of $m_1$ is (with {\boldmath $r$}$''$ = {\boldmath $r$}$'$)

\begin{eqnarray} \mbox{\boldmath $a$}_{{\star}o}& =&
\left \{ \frac{Gm_1 \mbox
{\boldmath $r$}'}{|\mbox{\boldmath $r$}'|^3}+Gm_2 \left [
\frac{\mbox{\boldmath $r$}'-
\mbox{\boldmath $r$}_{21}}{|\mbox{\boldmath $r$}'-
\mbox{\boldmath $r$}_{21}|^3}+
\frac{\mbox{\boldmath $r$}_{21}}{|\mbox{\boldmath $r$}_{21}|^3}
\right ] \right \}_{t=0}+ \nonumber\\
&& +{\beta}^2_0 \mbox{\boldmath ${\Omega}$}_0 \times \left (
\mbox{\boldmath ${\Omega}$}_0 \times
\mbox{\boldmath $r$}'_0 \right ). \label{aeso} \end{eqnarray}

The initial position {\boldmath $r$}$_0'$ of a surface element is
obtained with an initial spherical shape of star 1, and its initial
velocity is
{\boldmath $\vv$}$_0'$ = $({\beta}_0 - 1)$$\bf {\Omega}$$_0$ $\times$
{\boldmath $r$}$_0'$.

The contribution of the $m_2$- term in Eq. (\ref{aeso}) makes
{\boldmath $a$}$_{{\star}o}$ in a given parallel have a non-constant
azimuthal component. Thus the assumed initial spherical shape would
not be the proper shape consistent with this azimuthal behavior. In our
computations, at $t = 0$ we impose only the radial equilibrium
resulting from Eq. (\ref{aeso}), and allow unbalanced forces in the
azimuthal direction. We  found that with appropriate parameters 
(e.g. viscosity) this non-equilibrium initial condition will be
followed by a transient phase and a steady state (dependent on the
orbital phase) at later times.

At $t > 0$ the acceleration produced by gas pressure is conveniently
written in terms of the radial component of
{\boldmath $a$}$_{{\star}o}$. Thus the radial component in Eq.
(\ref{aeso}), which is used in the following section, is

\begin{eqnarray} 
a_{{\star}or'} &= &\frac{Gm_1}{{r'_0}^2}+   
  Gm_2 \left [ \frac{r'_0-{r_{21}}_0\sin{\theta}'\cos{\varphi}'_0}
{\left ({r'_0}^2+{r^2_{21}}_0-2r'_0{r_{21}}_0\sin{\theta}'
\cos{\varphi}'_0 \right )^{3/2}}\right] + \nonumber\\
&+& Gm_2 \left [\frac{\sin{\theta}'\cos{\varphi}'_0}{{r^2_{21}}_0}\right ]-
 {\beta}_0^2{\Omega}_0^2r'_0\sin^2{\theta}',
\label{aesor} \end{eqnarray}

\noindent with ${\theta}'$ the polar angle of the given parallel and
${\varphi}'_0$ the initial azimuthal angle of the element.

\subsection{The acceleration {\boldmath $a$}$_{\star}$}\label{aes}

An element on the stellar surface of $m_1$ has lateral surfaces facing  
in the three directions of spherical coordinates, $r'$, ${\varphi}'$,
${\theta}'$ in the primed non-inertial frame. The $x'$- axis points
to $m_2$ and ${\varphi}'$ = 0 on the positive side of this axis; the
polar axis is $z'$, which is also the stellar rotation axis. 
We will denote with $i$ the parallel's number and with $j$ the number
of the element. The lengths of an element in the three directions are
$l_{r'_{ij}}$, $l_{{\varphi}'_{ij}}$, and $l_{{\theta}'_{ij}}$, with
initial values $l_{r'_{ijo}}$, $l_{{\varphi}'_{ijo}}$,
$l_{{\theta}'_{ijo}}$. The length $l_{{\theta}'_{ij}}$ is assumed constant 
in time for all  elements in a given parallel. In our scheme, for a given element
 at any time $l_{{\varphi}'_{ij}}$ is the mean of its
azimuthal distances to the centers of mass of the two adjacent elements
in this azimuthal direction, and $l_{r'_{ij}}$ is twice the distance
between the center of mass of the element and the boundary of the
inner stellar region. At $t = 0$ all the elements in a given parallel
have the same lengths.

The acceleration {\boldmath $a$}$_{\star}$ has contributions from 
gas pressure and viscous shear. Below we describe  the
components of these contributions in spherical coordinates.

\subsubsection{Gas pressure}\label{pgas}

The gas pressure inside a surface element is $p_{ij}$, and we assume
a polytropic state equation $p_{ij} = p_{ijo}({\rho}_{ij}/{\rho}_{ijo})
^{{\gamma}'}$, with ${\gamma}' = 1 + 1/n$; $n$ is the polytropic index 
and $\rho$ the mass density. The pressure on the element exerted by  the
neighboring stellar inner region is taken as $p_{int} = p_{ij}/q$,
$0 < q < 1$. We consider in particular the half value $q = 0.5$. Thus 
the initial radial equilibrium is
$p_{ijo}l_{{\theta}'_{ijo}}l_{{\varphi}'_{ijo}}/q =
m_{ij}(a_{{\star}or'})_{ij}$, with $m_{ij}$ the constant mass of the
element. At $t > 0$ the radial pressure acceleration is
$(a_{{\star}r'1})_{ij} = p_{ij}l_{{\theta}'_{ij}}l_{{\varphi}'_{ij}}/
qm_{ij}$. This relation combined with the polytropic equation, the
initial radial equilibrium, and with the constant value of
$l_{{\theta}'_{ij}}$, gives

\begin{equation} (a_{{\star}r'1})_{ij} = \left ( \frac{l_{r'_{ijo}}}
{l_{r'_{ij}}} \right )^{{\gamma}'} \left ( \frac{l_{{\varphi}'_{ijo}}}
{l_{{\varphi}'_{ij}}} \right )^{{\gamma}'-1}(a_{{\star}or'})_{ij}. 
\label{aesr1} \end{equation}

The azimuthal gas pressure acceleration on a given element,
$(a_{{\star}{\varphi}'1})_{ij}$, is computed with the difference of 
the gas pressure forces exerted by the two adjacent elements in this
azimuthal direction. 
Then $(a_{{\star}{\varphi}'1})_{ij}  =
(p_{i,j-1}-p_{i,j+1})l_{r'_{ij}}l_{{\theta}'_{ij}}/m_{ij}$.
With the polytropic form for each pressure, the corresponding initial 
radial equilibrium conditions on the adjacent elements, and the same 
initial mass density for each surface element (thus the same mass of
elements in the given parallel, because their initial lenghts are the
same), this expression is

\begin{eqnarray} (a_{{\star}{\varphi}'1})_{ij}& =&
q \frac{l_{r'_{ij}}}{l_{{\varphi}'_{ijo}}} 
\left [ l_{r'_{ijo}}l_{{\varphi}'_{ijo}} \right ]^{{\gamma}'} \times \nonumber \\
&\times& \left \{(a_{{\star}or'})_{i,j-1}\left [ l_{r'_{i,j-1}}l_{{\varphi}'_{i,j-1}}
\right ]^{-{\gamma}'}   - (a_{{\star}or'})_{i,j+1}\left [
l_{r'_{i,j+1}}l_{{\varphi}'_{i,j+1}} \right ]^{-{\gamma}'} \right \}.\nonumber\\
&&  
\label{aesfi1}  \end{eqnarray}

\subsubsection{Viscous force}\label{fvis}

The shear part of the kinematic stress tensor is given by
(e.g., Symon 1971)

\begin{equation} {\bf P_{\eta}} = -{\eta} \left [ {\nabla}'
\mbox{\boldmath $\vv$}' + \left ( {\nabla}' \mbox{\boldmath $\vv$}'
\right )^{trp} - \frac{2}{3} {\bf 1} \left ({\nabla}' \cdot
\mbox{\boldmath $\vv$}' \right ) \right ],
\label{tensor} \end{equation}

\noindent where $\bf 1$ is the unit matrix, $trp$ indicates the
transposed, and ${\eta}$ is the coefficient of dynamic viscosity,
related with the coefficient of kinematic viscosity, ${\nu}$, by
${\eta} = {\nu}{\rho}$, with ${\rho}$ the mass density.
The total force on a given surface element from shear stresses is
$\bf F_{\eta}$ = $- \oint {\bf P_{\eta}} \cdot d{\bf S}$, with the 
integration over its surface. We picture each element with lateral
faces in the three directions of spherical coordinates, and
corresponding lateral areas $A_{r'}$, $A_{{\varphi}'}$,
$A_{{\theta}'}$. Because we do not compute the motion in the polar 
direction,  with ${\omega}' =
\vv'_{{\varphi}'}/r'\sin{\theta}'$, the relevant radial and azimuthal
components of the viscous force are

\[ (F_{\eta})_{r'}  =  A_{r'}{\Delta}\left [\eta \left (
\frac{4}{3}r'\frac{\partial}{\partial r'}\left ( \frac{\vv'_{r'}}{r'}
\right )-\frac{2}{3}\frac{\partial {\omega}'}{\partial {\varphi}'}
\right ) \right ] + \]  
\[+ A_{{\varphi}'}{\Delta} \left [\eta \left (
r' \frac{\partial {\omega}'}{\partial r'}\sin{\theta}'+
\frac{1}{r'\sin{\theta}'}\frac {\partial \vv'_{r'}}{\partial {\varphi}'}
\right ) \right ] + \]
\begin{equation}  + A_{{\theta}'}{\Delta} \left [
\frac{\eta}{r'} \frac{\partial \vv'_{r'}}{\partial {\theta}'} \right ],
\label{frvis}
\end{equation} 

\[ (F_{\eta})_{{\varphi}'}  =  A_{r'}{\Delta}\left [\eta \left (
r'\frac{\partial {\omega}'}{\partial r'}\sin{\theta}'+
\frac{1}{r'\sin{\theta}'}\frac {\partial \vv'_{r'}}{\partial {\varphi}'}
\right ) \right ]+ \]
\[+ A_{{\varphi}'}{\Delta} \left [\eta \left (
-\frac{2}{3}r' \frac {\partial}{\partial r'}\left ( \frac{\vv'_{r'}}
{r'} \right ) + \frac{4}{3}\frac{\partial {\omega}'}{\partial
{\varphi}'} \right ) \right ] + \]
\begin{equation} + A_{{\theta}'}{\Delta} \left [\eta 
\frac{\partial {\omega}'}{\partial {\theta}'}\sin{\theta}' \right ],
\label{ffivis}
\end{equation}

\noindent  with $\Delta$ meaning the difference of values of the
function inside a square parenthesis computed at the corresponding
lateral faces in the direction given by the lateral area factor. 
 
In our macroscopic element description, for the evaluation of the
several terms in this surface integration we consider as important the
gradients $across$ the lateral faces of an element and ignore local
gradients $along$ these faces. Thus, given that the density drops to
zero at the outer face and the radial velocity is also zero at the
inner rigid region, we approximate the radial and azimuthal
components of the viscous force as

\begin{equation} (F_{\eta})_{r'}  \simeq A_{{\varphi}'}{\Delta} \left [
\frac{\eta}{r'\sin{\theta}'}\frac {\partial \vv'_{r'}}{\partial
{\varphi}'} \right ] + A_{{\theta}'}{\Delta} \left [\frac{\eta}{r'}
\frac{\partial \vv'_{r'}}{\partial {\theta}'} \right ],
\label{apfrvis}
\end{equation}

\begin{eqnarray} (F_{\eta})_{{\varphi}'} &\simeq & -A_{r'}\left [\eta 
r'\frac{\partial {\omega}'}{\partial r'}\sin{\theta}' \right ] +
\frac{4}{3}A_{{\varphi}'}{\Delta} \left [\eta 
\frac{\partial {\omega}'}{\partial {\varphi}'} \right ] + \nonumber\\
&& +A_{{\theta}'}{\Delta} \left [\eta
\frac{\partial {\omega}'}{\partial {\theta}'}\sin{\theta}' \right ],
\label{apffivis}
\end{eqnarray}

\noindent with the first term in Eq. (\ref{apffivis}) computed at the
boundary with the inner rigid region.

The terms in Eq. (\ref{apfrvis}) give second and third contributions to
the radial acceleration, besides that in Eq. (\ref{aesr1}). Taking
the coefficient of kinematic viscosity, ${\nu}$, to be the same in all the
elements, the acceleration corresponding to the first term is
approximated as in Eq. (11) of Paper II

\begin{equation} (a_{{\star}r'2})_{ij} \simeq \frac{\nu}
{l^2_{{\varphi}'_{ij}}} \left [
\vv'_{{r'}_{i,j+1}} + \vv'_{{r'}_{i,j-1}} -2 \vv'_{{r'}_{ij}} \right ],
\label{aesr2} \end{equation}

\noindent likewise, the second term in Eq. (\ref{apfrvis}) gives the
acceleration

\begin{equation} (a_{{\star}r'3})_{ij} \simeq \frac{\nu}
{l^2_{{\theta}'_{ij}}} \left [
\vv'^{\ast}_{{r'}_{i+1,j}} + \vv'^{\ast}_{{r'}_{i-1,j}} -2 \vv'_{{r'}_{ij}}
\right ],
\label{aesr3} \end{equation}

\noindent with $\vv'^{\ast}_{{r'}_{i-1,j}}$, $\vv'^{\ast}_{{r'}_{i+1,j}}$
the mean radial velocities of the elements in the polar direction
adjacent to the element $ij$.

In the first term of Eq. (\ref{apffivis}) we approximate
$(\partial {\omega}'/\partial r') \simeq ({\omega}'-{\omega}'_{\star}) 
/l_{r'}$, with ${\omega}'_{\star} = {\beta}_0{\Omega}_0-\Omega$ the 
angular velocity of the inner stellar region as measured in the primed
non-inertial frame. In the second term we make a similar
approximation as that used to obtain Eq. (\ref{aesr2}) and ignore
the factor 4/3. Thus, the corresponding accelerations are

\begin{equation} (a_{{\star}{\varphi}'2})_{ij} \simeq - \frac{\nu}
{l^2_{r'_{ij}}} ( r'_{ij}- \frac{1}{2}l_{r'_{ij}}) \left [ \frac
{\vv'_{{\varphi}'_{ij}}}{r'_{ij}} - \left ( {\beta}_0{\Omega}_0 -
\Omega \right )\sin{\theta}'_i \right ].
\label{aesfi2} \end{equation}

\begin{equation} (a_{{\star}{\varphi}'3})_{ij} \simeq \frac{\nu}
{l^2_{{\varphi}'_{ij}}} \left [
\vv'_{{\varphi}'_{i,j+1}} + \vv'_{{\varphi}'_{i,j-1}} -
2 \vv'_{{\varphi}'_{ij}} \right ],
\label{aesfi3} \end{equation}

\noindent and finally, the last term in Eq. (\ref{apffivis}) gives the
azimuthal acceleration

\begin{equation} (a_{{\star}{\varphi}'4})_{ij} \simeq \frac
{{\nu}r'_{ij}}{l^2_{{\theta}'_{ij}}} \left [
({\omega}'^{\ast}_{i+1,j}-{\omega}'_{ij})\sin{\theta}'_{i,i+1}-
({\omega}'_{ij}-{\omega}'^{\ast}_{i-1,j})\sin{\theta}'_{i-1,i}
\right ], \label{aesfi4} \end{equation}

\noindent with ${\omega}'^{\ast}_{i-1,j}$, ${\omega}'^{\ast}_{i+1,j}$
the mean angular velocities of the elements in the polar direction
adjacent to the element $ij$, and ${\theta}'_{i-1,i}$,
${\theta}'_{i,i+1}$ the polar angles of the boundaries of the adjacent
parallels.

\subsection{Equations of motion}\label{ecsmov}

The acceleration {\boldmath $a$}$_{\star}$ of a surface element produced by 
gas pressure and viscous forces has a radial component
$a_{{\star}r'}$ given by the sum of Eqs. (\ref{aesr1}), (\ref{aesr2}),
and (\ref{aesr3}), and the corresponding azimuthal component
$a_{{\star}{\varphi}'}$ is the sum of Eqs. (\ref{aesfi1}),
(\ref{aesfi2}), (\ref{aesfi3}), and (\ref{aesfi4}). In the primed
non-inertial reference frame, the pair of radial and azimuthal
equations of motion of a surface element are

\begin{small}
\begin{eqnarray} \ddot{r}' & = & - \frac{Gm_1}{{r'}^2}-
Gm_2 \left [ \frac{r'-r_{21}\sin{\theta}'\cos{\varphi}'}
{\left ({r'}^2+r^2_{21}-2r'r_{21}\sin{\theta}'\cos{\varphi}'
\right )^{3/2}}+ \frac{\sin{\theta}'\cos{\varphi}'}{r^2_{21}}
\right ] + \nonumber \\
& & + a_{{\star}r'}+(\Omega + \dot{\varphi}')^2r'\sin^2{\theta}',
\label{emr} \end{eqnarray}

\begin{eqnarray} \ddot{\varphi}' & = & \frac{1}{r'\sin{\theta}'} 
\left \{ a_{{\star}{\varphi}'}- 
Gm_2 \left [ \frac{r_{21}}
{\left ({r'}^2+r^2_{21}-2r'r_{21}\sin{\theta}'\cos{\varphi}'
\right )^{3/2}}- \frac{1}{r^2_{21}} \right ]\sin{\varphi}'
\right \} - \nonumber \\
& & -\dot{\Omega} - \frac{2}{r'}(\Omega + \dot{\varphi}')\dot{r}',
\label{emfi}\nonumber \\
&& \end{eqnarray}
\end{small}

\noindent All  pairs of equations in all  surface elements are
solved simultaneously, along with the orbital motion of $m_2$ around
$m_1$. The values of $r_{21}$, $\Omega$, $\dot{\Omega}$ are obtained
from this orbital motion. We  used a seventh-order Runge-Kutta       
algorithm (Fehlberg 1968) to solve the equations. With some forty    
parallels covering the stellar surface, more than 10$^4$ elements are
employed.

\subsection{Method for $\dot{E}$ calculation}\label{disip}

The rate of energy dissipation per unit volume can be expressed in
terms of the tensor in Eq. (\ref{tensor}), and is given by the matrix
product (e.g.,McQuarrie (1976))      

\begin{equation} \dot{E_V} = - {\bf P_{\eta}}:{\nabla}' \mbox{\boldmath
$\vv$}'.
\label{dis1} \end{equation}

\noindent which can be written as

\begin{equation} \dot{E_V} = -2\eta \left [\frac{1}{3} \left ({\nabla}'
\cdot \mbox{\boldmath $\vv$}' \right )^2 - \left ( {\nabla}'
\mbox{\boldmath $\vv$}' \right )_s : \left ( {\nabla}'
\mbox{\boldmath $\vv$}' \right )_s \right ],
\label{dis2} \end{equation}

\noindent where $({\nabla}'${\boldmath $\vv$}$')_s$ is the symmetric
tensor $({\nabla}'${\boldmath $\vv$}$')_s$ = $\frac{1}{2} [{\nabla}'$
{\boldmath$\vv$}$'$+ $({\nabla}'${\boldmath $\vv$}$')^{trp}]$.

The computations in our model show that the gradient in azimuthal
velocity dominates that in radial velocity. Thus, Eq. (\ref{dis2}) can
be approximated as

\begin{eqnarray} \dot{E_V}&\simeq&{\eta} \left \{ \frac{4}{3} \left (
\frac{1}{r'\sin{\theta}'} \frac{\partial \vv'_{{\varphi}'}}{\partial
{\varphi}'} \right )^2 +
\left (\frac{\partial \vv'_{{\varphi}'}}{\partial r'} -
\frac{\vv'_{{\varphi}'}}{r'}\right )^2 +\right\} \nonumber\\
&+& {\eta}\left \{\frac{1}{r'^2} \left
(\frac{\partial \vv'_{{\varphi}'}}{\partial{\theta}'} -
\frac {\vv'_{{\varphi}'}}{\tan{\theta}'} \right )^2 \right \}.
\label{dis3} \end{eqnarray}

\noindent Now with ${\omega}' = \vv'_{{\varphi}'}/r'\sin{\theta}'$,
it follows that

\begin{equation}
\frac{\partial \vv'_{{\varphi}'}}{\partial r'} -
\frac{\vv'_{{\varphi}'}}{r'}
 =  r' \frac{\partial {\omega}'}{\partial r'}\sin{\theta}',
\label{trans1} \end{equation}

\begin{equation}
\frac{\partial \vv'_{{\varphi}'}}{\partial {\varphi}'}
 =  r' \frac{\partial {\omega}'}{\partial {\varphi}'}\sin{\theta}',
\label{trans2} \end{equation}

\begin{equation}
\frac{\partial \vv'_{{\varphi}'}}{\partial {\theta}'} -
\frac{\vv'_{{\varphi}'}}{\tan{\theta}'}
 =  r' \frac{\partial {\omega}'}{\partial {\theta}'}\sin{\theta}',
\label{trans3} \end{equation}

\noindent and Eq. (\ref{dis3}) reduces to

\begin{equation} \dot{E_V} \simeq {\eta} \left \{ \frac{4}{3} \left (
\frac{\partial {\omega}'}{\partial {\varphi}'} \right )^2 +
\left [ r'^2 \left ( \frac{\partial {\omega}'}{\partial r'} \right )^2 
+ \left ( \frac{\partial {\omega}'}{\partial {\theta}'} \right )^2
\right ]{\sin}^2{\theta}' \right \}.
\label{dis4} \end{equation}

\noindent For an accretion disk (${\theta}' = {\pi}/2$, and
the primed reference frame is taken as inertial in this case)
with ${\omega}'$ a function only of the radial distance $r'$, and
$\vv'_{r'}$ small compared with $\vv'_{{\varphi}'}$, Eq. (\ref{dis4}) gives
$\dot{E_V} \simeq {\eta}r'^2(\frac{d{\omega}'}{dr'})^2$ (Lynden-Bell \& Pringle 1974);     
this result was used in equation (14) of Toledano et al. (2007),
multiplied by the volume of an element.

The $j$- surface element on the parallel with polar angle
${\theta}'_i$ has a volume ${\Delta}V_{ij} =
r'^2_{ij}{\Delta}r'_{ij}{\Delta}{\varphi}'_{ij}$
${\Delta}{\theta}'_i\sin{\theta}'_i = l_{r'_{ij}}l_{{\varphi}'_{ij}}
l_{{\theta}'_{ij}}$, with 
$l_{{\varphi}'_{ij}} = r'_{ij}\sin{\theta}'_i{\Delta}{\varphi}'_{ij}$
and $l_{{\theta}'_{ij}} = r'_{ij}{\Delta}{\theta}'_i$.
The rate of energy dissipation in the element is
${\dot{E}}_{ij}={\dot{E}}_{Vij}{\Delta}V_{ij}$, with ${\dot{E}}_{Vij}$ the evaluation
of Eq. (\ref{dis4}) at the position of the element.

As already done to approximate Eq. (\ref{apffivis}), we take

\begin{equation}
\left (\frac{\partial {\omega}'}{\partial r'} \right )_{ij}
 =  \frac{{\omega}'_{ij}-{\omega}'_{\star}}{l_{r'_{ij}}},
\label{par1} \end{equation}

\begin{equation}
\left (\frac{\partial {\omega}'}{\partial {\varphi}'} \right )_{ij}
 = \frac{\frac{1}{2} \left ({\omega}'_{i,j+1}-{\omega}'_{i,j-1} \right
 )}{l_{{\varphi}'_{ij}}/r'_{ij}\sin{\theta}'_i},
\label{par2} \end{equation}

\begin{equation}
\left (\frac{\partial {\omega}'}{\partial {\theta}'} \right )_{ij}
 = \frac{\frac{1}{2} \left ({\omega}'^{\ast}_{i+1,j}-
{\omega}'^{\ast}_{i-1,j} \right )}{l_{{\theta}'_{ij}}/r'_{ij}},
\label{par3} \end{equation}

Eqs. (\ref{par1})-(\ref{par3}) are inserted in Eq. (\ref{dis4}),
giving ${\dot{E}}_{Vij}$ which has three contributions
$({\dot{E}}_{Vij})_{r'}$, $({\dot{E}}_{Vij})_{{\varphi}'}$, and
$({\dot{E}}_{Vij})_{{\theta}'}$, corresponding to the three gradients of
angular velocity in spherical coordinates. The total rate of energy
dissipation in the surface layer is

\begin{equation} {\dot{E}} = \sum_{i,j}^{} \left [
\left ({\dot{E}}_{Vij} \right )_{r'} + \left ({\dot{E}}_{Vij} \right )_
{{\varphi}'} + \left ({\dot{E}}_{Vij} \right )_{{\theta}'} \right ]
{\Delta}V_{ij}.
\label{etot} \end{equation}
\noindent In our computations the major contribution to
${\dot{E}}$ comes from $({\dot{E}}_{Vij})_{r'}$.

\section{Energy dissipation rates}

\subsection{Sample $\dot{E}({\varphi}',{\theta}')$ calculation}
\label{disfitet}

We can compute $\dot{E}_{ij}$, the rate of energy dissipation in the
element $j$ in a given parallel $i$, as a function of the azimuthal
angle ${\varphi}'$ of the element. Also, it is of interest to compute 
$\dot{E}_i$, the total rate of energy dissipation in the parallel
$i$, as a function of the corresponding polar angle ${\theta}'$.
As an example, in this section we show results from our model  in the
particular polytropic case with $n$=1.5 and using a surface layer depth
${\Delta}R_1/R_1$ = 0.03, with corresponding average mass density 
$\rho$ = 8.66$\times$10$^{-7}$ gr cm$^{-3}$.  The other input parameters are  
$M_1$=5 M$_\odot$, $M_2$=4 M$_\odot$, $R_1$=3.2 R$_\odot$, $\nu$ = 0.003 
$R^2_{\odot}$ day$^{-1}$, an orbital period $P$ = 10 days ($a$ = 40.62 $R_{\odot}$), 
and with two different values ${\beta}_0$ = 1.2 and 2.
\begin{figure}
\centering
\includegraphics[width=0.80\linewidth]{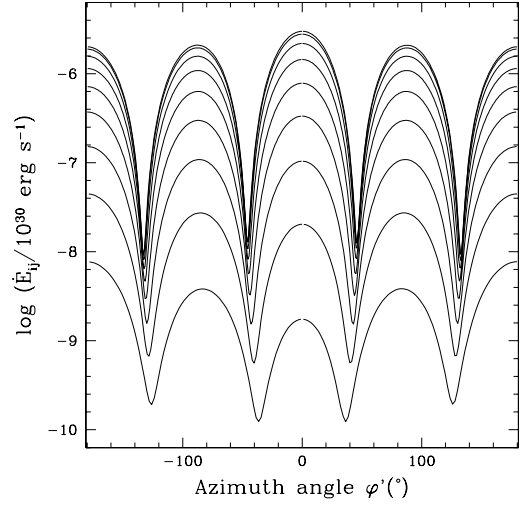}
\caption {Rate of energy dissipation $\dot{E}_{ij}$ as a function
of ${\varphi}'$ in a few selected parallels, for the n=1.5 polytropic case
${\Delta}R_1/R_1$ = 0.03, $\nu$ = 0.003 $R^2_{\odot}$ day$^{-1}$,
$\rho$ = 8.66$\times$10$^{-7}$ gr cm$^{-3}$, orbital period $P$ = 10 days
($a$ = 40.62 $R_{\odot}$), and ${\beta}_0$ = 1.2.  The top  curve shows results for
the stellar equator, and following curves from top to bottom correspond to parallels
with ${\theta}'$ $\approx$ 81$^\circ$, 73$^\circ$, 64$^\circ$, 55$^\circ$, 46$^\circ$,
38$^\circ$, 29$^\circ$, and 20$^\circ$.  Calculations
with ${\beta}_0$ = 2.0 yield identical curves, but the $\dot{E}_{ij}$ values are higher for
by $\sim$1.5 magnitudes.
}
\label{15874fg1}
\end{figure}

Figure \ref{15874fg1}  shows the run of $\dot{E}_{ij}$ as a function of
${\varphi}'$ in some selected parallels for  ${\beta}_0$=
1.2. The upper curve corresponds to the stellar equator, ${\theta}'$
= 90$^\circ$, and following curves from top to bottom give results for
the parallels with ${\theta}'$ $\approx$ 81$^\circ$, 73$^\circ$, 64$^\circ$,
55$^\circ$, 46$^\circ$, 38$^\circ$, 29$^\circ$ and 20$^\circ$.  
A very similar results is obtained for the ${\beta}_0$=2 calculation, although
the energy dissipation rates are significantly larger.

\begin{figure}
\centering
\includegraphics[width=0.80\linewidth]{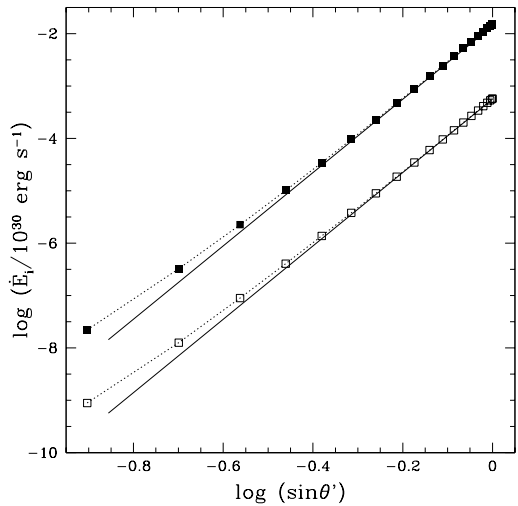}
\caption {Rate of energy dissipation in the parallel $i$ as a function
of its polar angle, corresponding to the computations in Figure 1.
Results with ${\beta}_0$ = 1.2 and 2 are
shown with empty and filled squares, respectively. The continuous
lines show functions with a $\sin^7{\theta}'$ dependence.}
\label{15874fg2}
\end{figure}

Figure \ref{15874fg2} shows the total rate of energy dissipation in the
parallel $i$, $\dot{E}_i$, as a function of its polar angle.
Both cases ${\beta}_0$ = 1.2 and 2 are shown in this figure, with
empty and filled squares, respectively. The continuous lines show
functions with a $\sin^7{\theta}'$ dependence. Thus, our computations
give the approximate dependence $\dot{E}_i \sim \sin^7{\theta}'_i$,
excluding polar angles in a region of approximately 10$^\circ$
around the poles. This implies that in our model
$(\frac{\partial {\omega}'}{\partial r'})_i$ has a $\sin^2{\theta}'_i$
dependence, thus the dominant radial gradient in Eq. (\ref{dis4})
integrated in the volume of the parallel, which has a $\sin{\theta}'_i$
dependence, gives $\dot{E}_i \sim \sin^7{\theta}'_i$. Concerning this
result, there is a theoretical study by Scharlemann (1981), which gives
the tidal velocity field in a differentially rotating convective
envelope of a component in a binary system in circular relative orbit,
in the limit ${\beta}_0 \rightarrow 1$. For the case in which the
star rotates uniformly, equation (40c) of Scharlemann (1981) gives the
azimuthal component of the tidal velocity field. This equation
implies, in our notation, $(\frac{\partial {\omega}'}{\partial r'})_i
\sim \sin^3{\theta}'$, which differs from our result. Yet 
 we find in our computations with ${\beta}_0$ = 1.2, 2.0 that the 
surface layer does not rotate uniformly, contrary to the assumption in
Scharlemann's equation.  Thus, the difference between the two results
most likely stems from Scharlemann's  assumption of uniform rotation 
over the stellar surface which, for $\beta\neq$1, is incorrect.

\subsection{Dependence of  $\dot{E}$ on orbital separation}

In this section, we analyze the behavior of the total rate of energy
dissipation, $\dot{E}$ for  binary system models with different
orbital separations and  a variety of orbital eccentricities.

\subsubsection{Circular orbits} \label{circsect}

 As in Toledano et al. (2007), we
adopt as the test binary system one with masses $m_1$ = 5 $M_{\odot}$ and
$m_2$ = 4 $M_{\odot}$; $m_1$ with a radius $R_1$ = 3.2 $R_{\odot}$.
The analysis was made for  circular relative orbits  with several different values of
orbital separation $a$. The energy dissipation rate given by Eq. (\ref{etot}) was 
computed in $m_1$; we used 20 parallels distributed between the equator and polar 
angle 85$^\circ$, with more than 10$^4$ surface elements, and a polytropic index n=1.5. In the 
coefficient of dynamic viscosity ${\eta} = {\nu}{\rho}$  the same approach as 
that of  Toledano et al. (2007) was adopted.  That is,  the average mass density for the 
surface layer, $\Delta R_1/R_1$, is taken from  a  BEC  stellar structure model 
computation.\footnote{The Binary Evolutionary Code, Langer (1991)} For this paper
we used a model\footnote{$m_1$= 5M$_\odot$, with an age of 4.29125e7 yrs, at which 
time its radius is 3.156 R$_\odot$.}  kindly 
provided by I. Brott (private communication, 2010).
The value of the kinematic viscosity $\nu$ was taken as the  lowest 
value allowed by the code (see discussion in Harrington et al. (2009) regarding this 
parameter).  For the shortest period binary models with circular orbits, this value is 
$\nu$ = 0.003 $R^2_{\odot}$ day$^{-1}$, which we adopted for the whole set of circular 
orbit  models discussed in this section. Three different values of the thickness of 
the surface layer in $m_1$ were considered, ${\Delta}R_1/R_1$ = 0.01, 0.03, 0.06.\footnote{Appendix 1
shows the manner in which the $\dot{E}$  results depend on the chosen depth of the surface layer.}   
The corresponding average mass densities are,  1.19$\times$10$^{-8}$, 1.68$\times$10$^{-7}$ 
and 8.66$\times$10$^{-7}$ gr cm$^{-3}$, respectively. The computations were made for 
two values ${\beta}_0=$ 1.2, 2.0  of the synchronicity parameter.


\begin{figure}
\centering
\includegraphics[width=0.80\linewidth]{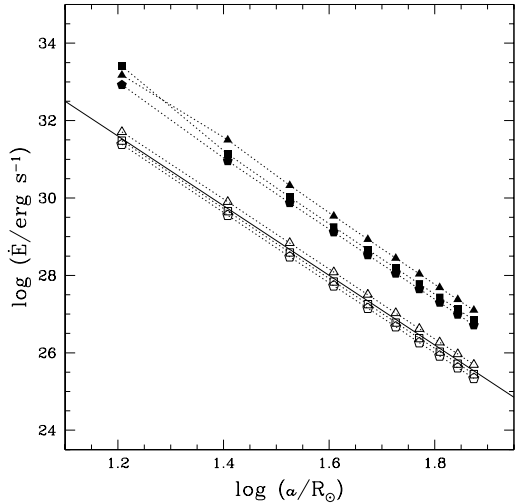}
\caption {Energy dissipation in $m_1$ as a function of the
separation $a$ of the circular orbit. A polytropic index $n$ = 1.5 is used.
Empty symbols correspond to ${\beta}_0$ = 1.2 and filled symbols to
${\beta}_0$ = 2.0. The correspondence with the layer thickness
${\Delta}R_1/R_1$ is, in triangles: ${\Delta}R_1/R_1$ = 0.01, in squares:
${\Delta}R_1/R_1$ = 0.03, in pentagons: ${\Delta}R_1/R_1$ = 0.06. The
continuous line is a function with an $a^{-9}$ dependence on $a$.}
\label{15874fg3}       
\end{figure}

Figure \ref{15874fg3} shows the energy dissipation ${\dot{E}}$ in $m_1$ as a function of the 
radius of the circular orbit. Empty symbols
correspond to ${\beta}_0$ = 1.2 and filled symbols to ${\beta}_0$ =
2.0. Triangles, squares, and pentagons show results for  ${\Delta}R_1/R_1$ = 0.01,
0.03, and 0.06, respectively. A clear linear behavior is obtained on this
log-log plot. For comparison, the continuous line in this
figure shows a function with an $a^{-9}$ dependence on the orbital
radius $a$. The data in the three cases ${\Delta}R_1/R_1$ would shift vertically
for different values of the mass density and viscosity, but the functional
dependence on orbital separation remains the same.

\subsubsection{Elliptic orbits}\label{ellip}

Elliptical orbit binary systems differ from those in circular orbits
in that the synchronicity parameter, $\beta$, does not remain constant
over the orbital cycle.  This is because of the variation of the orbital angular velocity,
$\Omega$.  The computation takes into account the changing value of $\beta$, but we
characterize each model by its synchronicity parameter at periastron, $\beta_0$.

\begin{figure}
\centering
\includegraphics[width=0.80\linewidth]{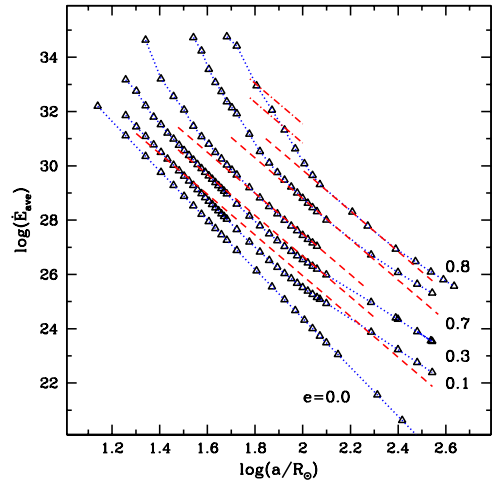}
\caption{Energy dissipation rates (in ergs s$^{-1}$) from our model calculations for $\beta_0=$1.2
and different eccentricities as indicated.  The  dashed lines show the absolute value of
$\dot{E}_{orb}$ given by  Eq. A31 of Hut (1981), as written in  our Eq. (\ref{eqhut2}).
The dotted lines connect the points corresponding to each eccentricity. The short dot-dashed lines
show the values of $\dot{E}_{Hut}$ for $e$=0.8 and lag angles $\alpha$=10$^\circ$ and 50$^\circ$.}
\label{15874fg4}             
\end{figure}

The energy dissipation rates for a grid of elliptical orbit binary systems
were computed following a similar procedure as in Section \ref{circsect} for the 
$m_1+m_2=$ 5$+$4 M$_\odot$ binary system, but in this case eccentricities 
$e=$0.1, 0.3, 0.5 0.7 and 0.8 were assigned, and 20 latitude grid points were used
in each hemisphere of $m_1$.  
The remaining parameters were held constant:
$R_1=$3.2, $\beta_0=$1.2, $\Delta R_1/R_1=$0.06, $n=$1.5, $\nu=$0.005 R$^2_\odot/day$.  The changing 
separation of the two stars over the orbital cycle leads to a strong dependence of the energy 
dissipation rate on the orbital phase. Thus, in order to assign a value of $\dot{E}$ to each
model calculation, an average value of the energy dissipation rate over orbital cycle,
$\dot{E}_{ave}$, was computed by numerically integrating $\dot{E}$ obtained at many phases 
distributed over the orbital cycle, and then dividing by the corresponding orbital period. 
The values of $\dot{E}_{ave}$ are plotted in Figure \ref{15874fg4} as a function of the major 
semi-axis of the orbit.             

Two aspects of this figure stand out: 1) there is a ``family" of curves, one for each value of
the eccentricity; 2) in each case, the scaling of $\dot{E}_{ave}$ with $a$ does not follow 
a unique linear (in the log-log plane) plot, as in the circular orbit case.  Instead,
for very short orbital periods, the slope is steeper and for very long periods, the slope is flatter
than the $\dot{E}_{ave}\sim a^{-9}$ relation for circular orbits.  It is also interesting to note 
that the departure from this relation grows with increasing orbital eccentricity.  

\subsection{Comparison with the ``weak friction" model}

Hut (1981) provided a derivation of the differential equations for the tidal evolution  
under the ``weak friction, equilibrium tide" model representation for the tidal interaction 
(Darwin 1880; Alexander 1973).   
The fundamental assumption of this model is that  the tidal bulges that are raised by the external 
perturbing potential are only slightly misaligned with respect to the axis that connects the two stars. The system is then
driven toward an equilibrium configuration  through the  action of the torques acting on the ``retarded
body". The misalignment of the bulges is caused by dissipation (i.e., friction) in the perturbed
star. Hut's formalism includes high order terms in the orbital eccentricity, as required to properly
assess the energy dissipation rates in very eccentric systems.\footnote{The model has recently been 
revisited by Leconte et al. (2010) in the context of tidal heating in exoplanets. These authors  
extended Hut's equations to the case in which the orbital plane and the rotational equatorial 
plane of the objects do not coincide.  A  similar extension was also given in Eggleton et
al. (1998).}
In this section we test  our model by computing the energy dissipation rates for a  5+4 M$_\odot$  
polytropic (n=1.5) binary system with a primary star's radius $R_1$=3.2 R$_\odot$ for a range in
orbital separations and eccentricities. Each binary system of the grid is characterized
by the energy dissipation rates averaged over the orbital cycle, $\dot{E}_{ave}$.  We show that
for a variety of orbital eccentricities, the behavior of $\dot{E}_{ave}$ is consistent with the
predicted energy dissipation rates in the ``weak friction" approximation given by Hut (1981) within
the domain of applicability of this approximation.  Our model also adequately predicts the
pseudo-synchronization angular velocity obtained by Hut (1981) for moderate eccentricities ($e\leq$0.3).

\subsubsection{Energy dissipation rates}

The rate of energy dissipation within the primary star associated with the decrease in the orbital and 
rotation energy in the system is given by Hut (1981) in his Eq. A31 

\begin{eqnarray} \dot{E}_{Hut}  =  -3 \frac{k}{T} G(m_1+m_2)\left ( \frac{m_2^2}{m_1} \right ) \left(\frac{R_1^8}{a^9}\right )  (1-e^2 )^{-15/2} \left [A-2Bx+Cx^2 \right ] \nonumber \end{eqnarray}


\noindent where $k$ is a constant that depends on  the  stellar structure, and $x=\omega/n$, with $\omega$  
the  angular rotation  velocity of $m_1$ and $n=n_p (1-e^2)^{3/2}/(1+e)^2$ the mean orbital angular velocity,  
with $n_p$ the orbital angular velocity  at periastron. $T=R_1^3/Gm_1\tau$, where $\tau$ is the lag time, and

\begin{eqnarray}
A&=&1+\frac{31}{2}e^2+\frac{255}{8}e^4+\frac{185}{16}e^6+\frac{25}{64}e^8,  \nonumber   \\
B&=&1+\frac{15}{2}e^2+ \frac{45}{8}e^4+ \frac{5}{16}e^6,  \nonumber \\
C&=&1 +3 e^2+\frac{3}{8}e^4. 
\end{eqnarray}

\noindent After converting masses and 
distances to solar units, and writing  $\omega/n$ in terms of $n_p$, and recalling 
that $\omega/n_p=\beta_0$, the above equation  can be rewritten as 

\begin{eqnarray}  
\dot{E}_{Hut}& =& -4.4\times10^{42} k \tau \left ( \frac{m_1+m_2}{M_\odot} \right )
\left (\frac{m_2}{M_\odot} \right)^2 \left (\frac{R_1}{R_\odot} \right )^5 \left(\frac{a}{R_\odot} \right)^{-9} \times \nonumber \\
&&\times  (1-e^2)^{-15/2}
 \left [A - 2 Bx+Cx^2 \right ],
\label{eqhut2} \end{eqnarray}

\noindent  with $\dot{E}_{Hut}$ given in ergs s$^{-1}$ and  $x=\beta_0 (1+e)^2$.

For the lag time, Hut (1981; Eq. 4) gives $\tau=\alpha/(\omega-\Omega)$, where $\omega$ and $\Omega$
are the rotational and instantaneous orbital angular velocity, respectively.  $\alpha$ is the lag 
angle, and is measured from the line that joins the centers of $m_1$ and $m_2$. This
expression may be re-written in terms of the  instantaneous synchronicity parameter, $\beta$, as
$\tau=\alpha/\Omega(\beta-1)$.  

In  eccentric binary
systems, $\alpha$ is generally very small around periastron, but it can be quite large at other
orbital phases.\footnote{An example of the orbital-phase dependence of $\alpha$ computed with our
model may be found in Koenigsberger et al. 2003.}  At the same time, the amplitude of the tidal
bulges is largest around periastron and, given that this is when $m_1$ and $m_2$ have their closest
approach, the tidal interaction effects are strongest.  Hence, for the purpose of these exploratory
calculations, we now make the following  approximation: To compute the energy dissipation rates
using Eq. (\ref{eqhut2}), we adopt $\tau\sim\alpha_{per}/\Omega_0(\beta_0-1)$, where $\alpha_{per}$  is
the lag angle at periastron which we assume is relatively small, and here tentatively adopt $\alpha$=1$^\circ$.
We thus also use $\beta=\beta_0$.  Using  Kepler's relation 
$\Omega_0=G^{1/2} (m_1+m_2)^{1/2} a^{-3/2} (1+e)^{2}(1-e^2)^{-3/2}$ and converting to solar units,

\begin{eqnarray}
\tau \sim 1.6\times10^3  \left(\frac{a}{R_\odot}\right)^{3/2} \left (\frac{m_1+m_2}{M_\odot}\right )^{-1/2} \frac{(1-e^2)^{3/2}}{(1+e)^2} 
\frac{\alpha_{per}}{(\beta_0-1)} 
\label{eqT} \end{eqnarray}

\noindent with $\alpha_{per}$ given in radians and $\tau$ is in seconds.  Note that when Eq. (\ref{eqT}) is 
introduced into Eq. (\ref{eqhut2}), the dependence of $\dot{E}_{Hut}$ on orbital separation $\sim a^{-15/2}$, 
thus flattening the slope from the $a^{-9}$ relation that holds for a circular orbit.

For the value of $k$, we use the expression
given in Lecar et al. (1976; their eq. (2)) for the mass in the tidal bulge, $M_t$:
\begin{eqnarray}
M_t=\frac{1}{2} k M_1 \frac{h}{R_1}
\label{k_lecar} \end{eqnarray}

\noindent where $h$ is the height of the tide.  Consider, for example, the model binary system
with $e$=0.8, $P$=15 d, from the grid of models discussed in the previous section.  The height
of the tide at periastron is $\sim$0.03 R$_\odot$ (see below).  The mass in the bulge is
$M_t = f_tM_{layer}$, where $f_t$ is the  fraction of the mass in the outer layer, $M_{layer}$, 
that goes into to the bulge.  For the particular case in question, 
$M_{layer}=4\pi R_1^3 \Delta R_1 <\rho>$=3.67$\times$10$^{-6}$ M$_\odot$, where $\Delta R_1$=0.06 R$_\odot$ 
is the layer thickness and $<\rho>$=8.66$\times$10$^{-7}$ g cm$^{-3}$, is the average  density of the layer.   
The maximum in the primary bulge  approximately extends over 20$^\circ$ in azimuth, so $f_t\sim$10$^{-4}$.
Hence, this  estimate yields $k\sim$10$^{-8}$.  

With these considerations, we use Eq. (\ref{eqhut2}) to estimate the energy dissipation rates in the
``weak friction" approximation, $\dot{E}_{Hut}$, for the binary model parameters $m_1$=5 M$_\odot$,
$m_2$=4 M$_\odot$, $R_1$=3.2 R$_\odot$,  $\beta_0$=1.2, and $k$=5$\times$10$^{-8}$.  The dashed lines in 
Figure \ref{15874fg4} correspond to the absolute values of  $\dot{E}_{Hut}$ given by Eq. (\ref{eqhut2}).   
The coincidence between the general behavior of these lines and that of the numerical computations is  
encouraging. Note that  no  relative shift or scaling is introduced, and yet the relative separation 
between the curves corresponding to the different eccentricities is reproduced by our models. In addition, 
our model reproduces the general trend in the slope of the $\dot{E}_{ave}$ {\it vs.} $a$ relation, particularly 
for large orbital separations.  Thus, we find  that our numerical calculation captures to a large extent the 
general scaling in the energy dissipation rates predicted analytically within the ``weak friction" 
tidal model. 

However,  although the scaling with $e$ predicted by Eq. (\ref{eqhut2}) is adequately 
captured by our model,  the  dependence on $a^{-n}$, with $n$=constant is not satisfied.  Specifically, there
is a marked excess in $\dot{E}_{ave}$ for smaller orbital separations, with respect to $\dot{E}_{Hut}$. 
Several factors may be 
responsible for this difference. First, the assumption of a constant lag angle, $\alpha$, is quite inadequate.  
For example,  assuming that for the $e$=0.8 calculation $\alpha$=10$^\circ$ and 50$^\circ$, instead 
of the 1$^\circ$ used to draw the dash lines,  significantly higher $\dot{E}_{Hut}$ values are 
obtained.\footnote{We note that at periastron, $\alpha$ is indeed very small (see, for example, 
Figure \ref{15874fg10}), but grows after periastron passage.}  This 
is illustrated by the short dashed lines in Figure \ref{15874fg4}.   
Second,  the applicability of the standard ``equilibrium tide" model for high  eccentricities
is questionable (Efroimsky \& Williams 2009).  Finally, we note that our model assumes  a spherical
rigidly-rotating interior region, and that  this approximation is less applicable  for small orbital
separations, where the deformation of the star is greater. 

\begin{figure}
\centering
\includegraphics[width=0.80\linewidth]{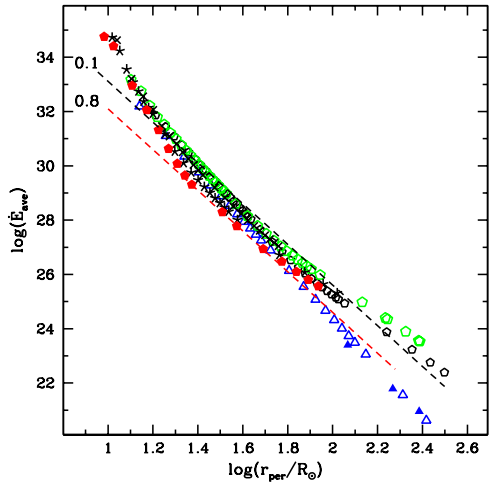}
\caption{Energy dissipation rates  of the previous figure but
here plotted as a function of the periastron distance, $r_{per}$. In
this representation, the different curves for each eccentricity
``collapse" onto a common region. The dashed lines show the absolute value
of $\dot{E}_{orb}$ given by Eq. A31 of Hut (1981) for $e$=0.1 and 0.8, also plotted
as a function of $r_{per}$. Symbols correspond to the different eccentricities:
$e$=0.00 (open triangles), 0.01 (filled triangles), 0.10 (small pentagons), 0.30
(large pentagons), 0.50 (crosses), 0.70 (stars) and 0.8 (filled pentagons).}
\label{15874fg5}         
\end{figure}

It is interesting to note that if  $\dot{E}$ is plotted as a function of $r_{per}$, the
orbital separation at periastron, instead of the semi-major axis,  the curves corresponding
to the different eccentricities ``collapse" toward a single curve.  This is illustrated in
Figure \ref{15874fg5}.  Thus,  the periastron distance is a more convenient 
parameter for the purpose of analyzing the average energy dissipation rates than the
semi-major axis.


\subsubsection{Pseudo-synchronization angular velocity}

Pseudo-synchronization in an eccentric binary is a term used to describe the state  that is
closest to an equilibrium configuration.   Hut (1981) demonstrated that the rotational angular velocity for
pseudo-synchronization, $\omega_{ps}<\Omega_0$, where $\Omega_0$ is the orbital angular velocity
at periastron. Specifically,  0.8$<\omega_{ps}/\Omega_0<1$.\footnote{Recall  that
in circular orbits, equilibrium configuration is at $\beta_0$=$\omega/\Omega_0$=1, where  $\dot{E}$=0.}

\begin{figure}
\centering
\includegraphics [width=0.45\linewidth]{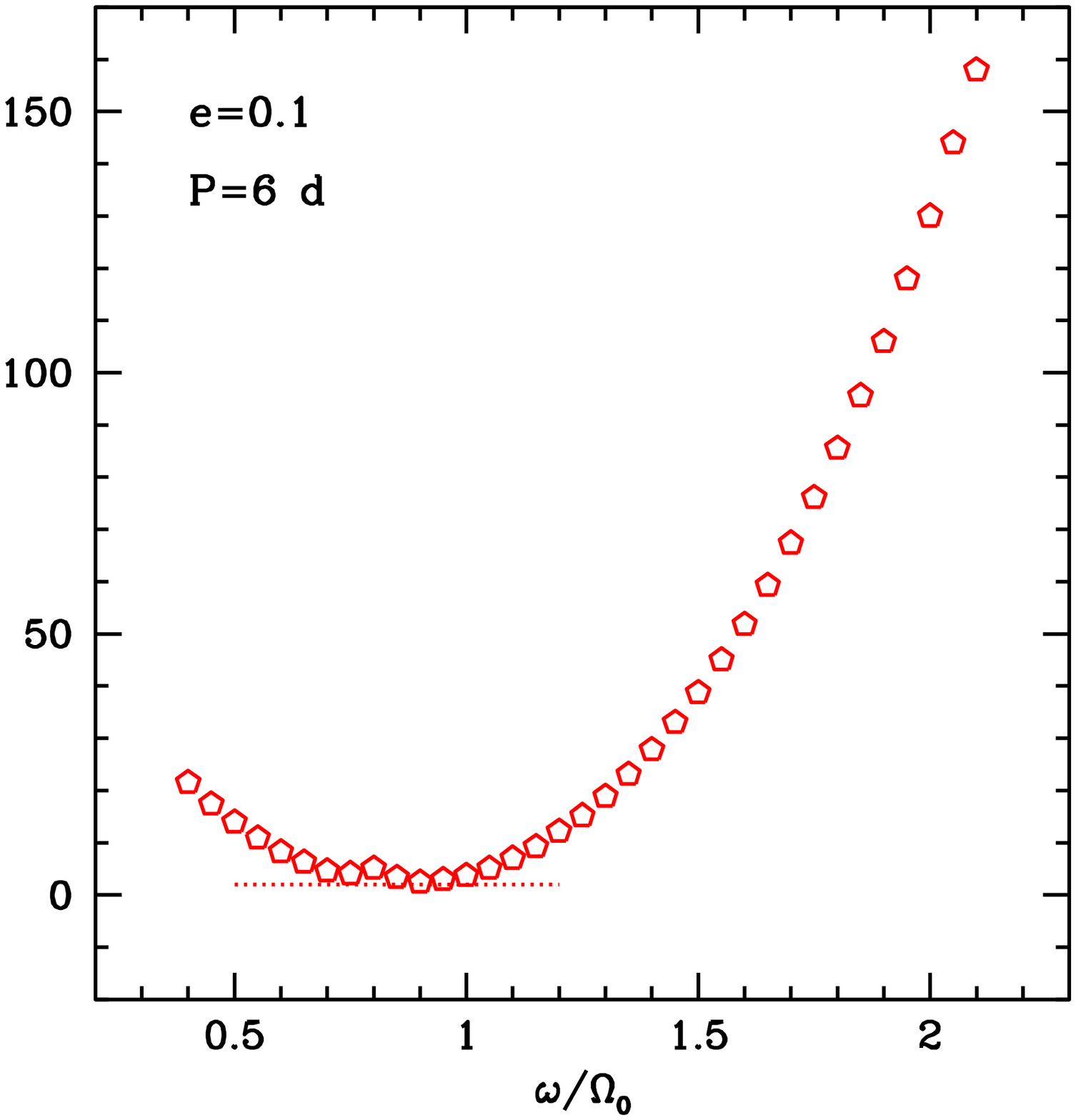}     
\includegraphics [width=0.45\linewidth]{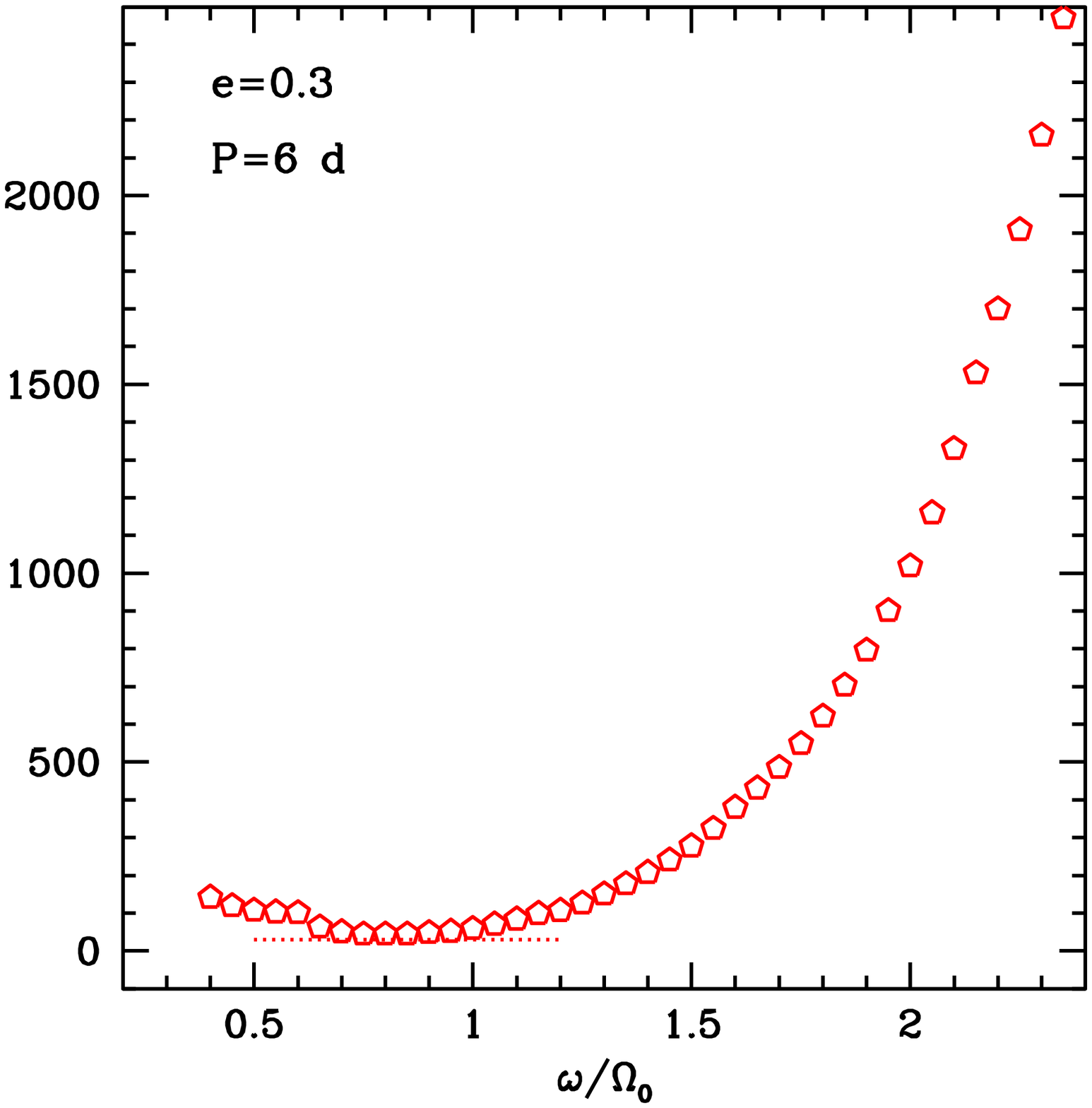}
\includegraphics [width=0.45\linewidth]{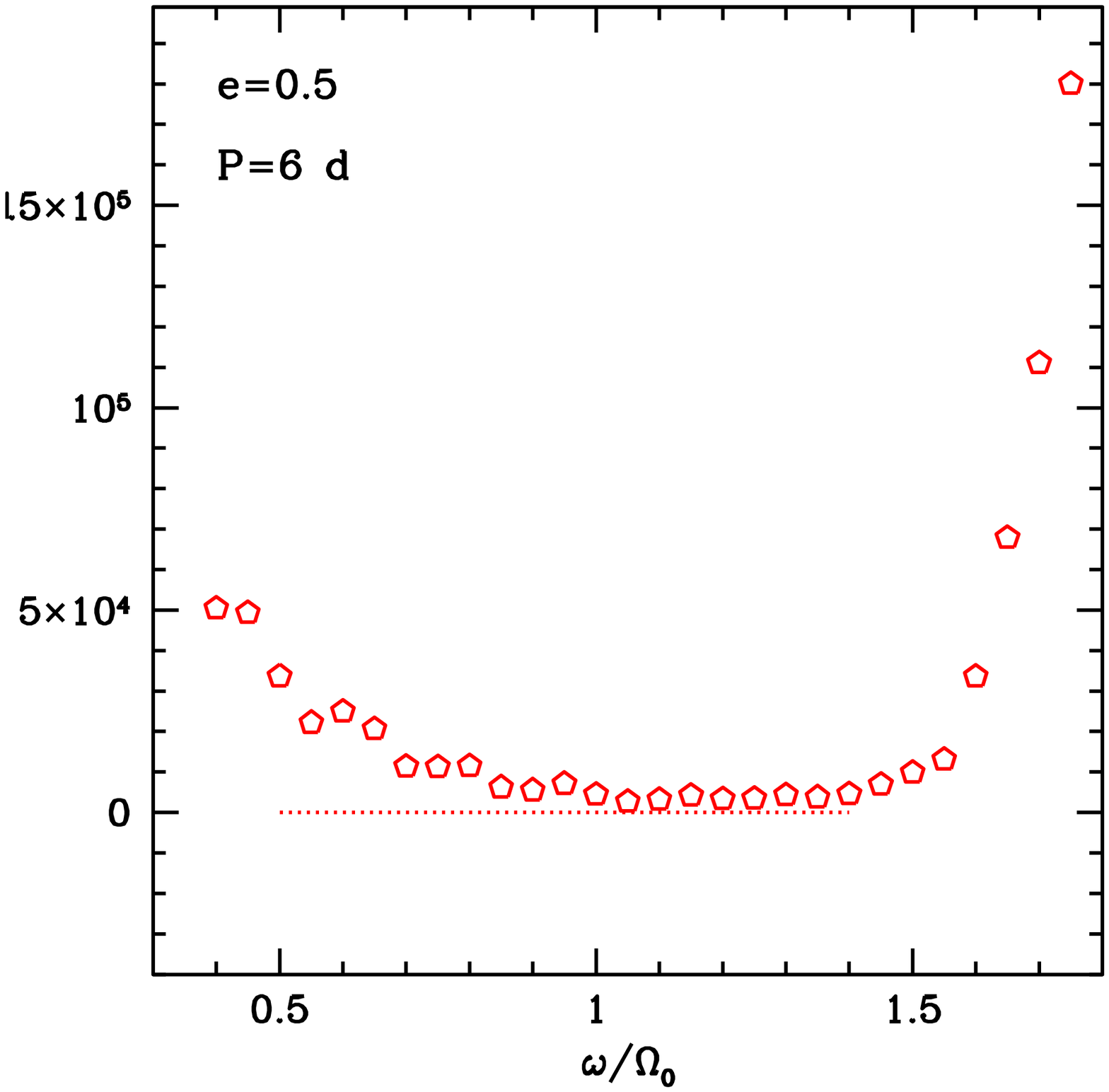}
\includegraphics [width=0.45\linewidth]{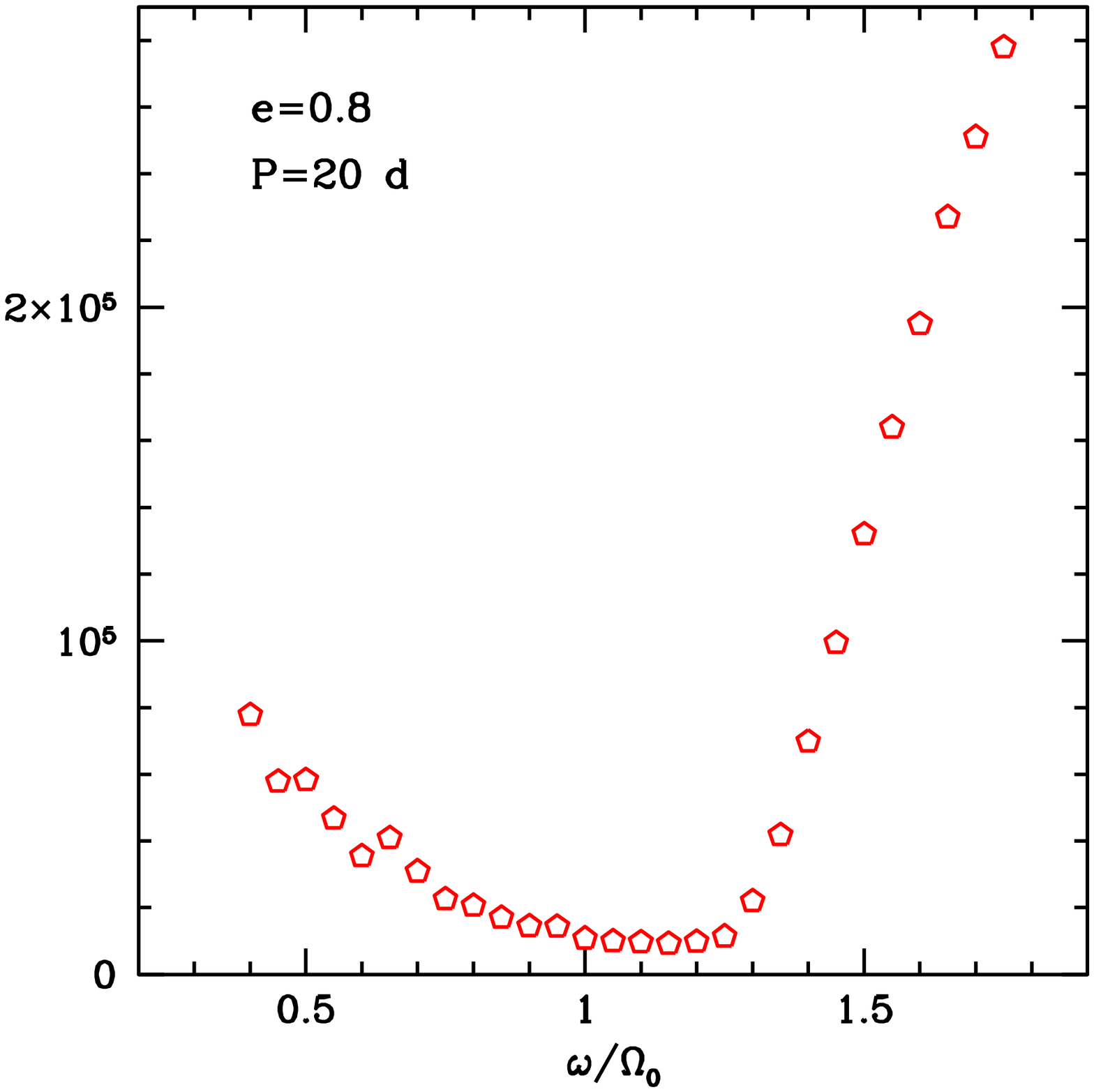}
\caption {Energy dissipation rates $\dot{E}_{ave}$/10$^{35}<\rho>$ as a function of
$\beta_0=\omega/\Omega_0$ for four different eccentricities, $e=$0.1 (top left), 0.3 (top right),
0.5 (bottom left) and 0.8 (bottom right).  The first three cases are for an orbital
period $P$=6 d and the last is for 20 d. The ordinate is in units of 10$^{35} <\rho>$ergs s$^{-1}$,
where $<\rho>$ is the average density.  The dotted line indicates the level of minimum $\dot{E}_{ave}$ which, for
$e$=0.1 and 0.3 coincides with the  analytical results of Hut (1981).
}
\label{15874fg6}   
\end{figure}

We computed sets of models with $e$=0.1, 0.3, 0.5 ($P$=6 d) and 0.8 ($P$=20 d) holding all other
parameters fixed except for $\beta_0$, which was varied from $\sim$0.4 to the highest value admitted 
by the calculation.\footnote{For the fixed value of $\nu$ chosen for these calculations, very 
high stellar rotation rates cause neighboring surface elements to overlap, a condition which 
halts the computation.}   Figure \ref{15874fg6} illustrates the  dependence of  $\dot{E}_{ave}$ on 
$\beta_0$.  In all cases, there is a minimum in $\dot{E}_{ave}$ at a particular value (or range of values) 
of $\beta_0$.  For e=0.1 and 0.3, the minimum occurs around $\beta_0 \sim$0.8--0.9, consistent with the 
``weak friction" approximation. For e=0.5 and 0.8, however, the minimum occurs for  1.0$<\beta_0<$1.4, a 
result that is likely related to the caveats already mentioned at the end of the previous section.

\section{Synchronization timescales}\label{tsincro}


In this section we analyze the synchronization time scales, which will be defined 
below, using the  model presented in Section \ref{modeloext}, the
procedure to compute the energy dissipation in Section \ref{disip}, and the
$\dot{E}_{ave}$ values described in the previous section.  The results
will be  compared with the earlier results of  Toledano et al. (2007) and with a 
theoretical result given by Zahn (1977, 2008). 

\subsection{Circular orbits}

As in Toledano et al. (2007), we now procede to estimate the synchronization time 
scale between the stellar rotation period and the orbital period. The  synchronization 
time scale is the time needed by the system to arrive at a state in which both periods 
are equal, and this  state is here assumed to be attained as a consequence of the 
energy dissipation in the surface layer. In our computations, at time $t = 0$ the whole star,
including the surface layer, is rotating with  an angular velocity
{\boldmath ${\omega}$}$_{\star}$ = ${\beta}_0$$\bf {\Omega}$$_0$,
with $\bf {\Omega}$$_0$ the reference orbital angular velocity.  For circular orbits, 
$\bf {\Omega}$$_0$ is directly the angular velocity in the circular orbit while for $e>$0 it is
the angular velocity at periastron. We adopt here as the condition for synchronicity\footnote{Note, 
however, as shown  by Hut (1981), the lowest energy state for the eccentric system corresponds to  
the pseudosynchronization period which, for moderate eccentricities, is ${\beta}\sim$0.8--0.9.}
${\beta}_0$ = 1. Thus, with ${\beta}_0 > 1$ there is an initial excess of rotational kinetic
energy over that in the synchronized state. As a first approximation,
ignoring the orbital change suffered by the binary system to reach the ${\beta}_0$ = 1
state,  the excess of rotational kinetic energy is proportional
to $({\beta}^2_0 -1){\bf {\Omega}}^2_0$, with the proportionality constant being
$I/2$, with $I$ the moment of inertia. 
Thus for star $m_1$ the synchronization time is on the order of

\begin{equation} {\tau}_{syn} \approx \frac{I({\beta}^2_0 -1){\bf {\Omega}}^2_0}{2{\dot{E}}}.
\label{tsin}
\end{equation}

From Kepler's law, ${\bf {\Omega}}^2_0$ has an $a^{-3}$ dependence on
the major semi-axis $a$. Thus, for a fixed value of ${\beta}_0$ and 
with the approximate $a^{-9}$ dependence of ${\dot{E}}$ from
Figure \ref{15874fg3}, ${\tau}_{syn}$ will have an $a^6$ dependence on $a$       
for circular orbits.

On the other hand, Zahn (1977) has analyzed theoretically the tidal
friction in close binary stars, also arriving at an $a^6$ dependence 
for stars in which a turbulent viscosity may be the main mechanism that produces the 
tidal friction. In a recent review of the topic (Zahn 2008), this expression  
for circular orbits is given as 

\begin{equation} 
\frac{1}{{\tau}_{Zahn}} = \frac{1}{t_{diss}} \frac{(\omega-\Omega)}{\omega} q^2
\frac{m_1R_1^2}{I}\left (\frac{R_1}{a}\right )^6,
\label{tZsin}
\end{equation}

\noindent with $q=m_2/m_1$, $I$ the moment of inertia of $m_1$, $\Omega$ is the orbital angular velocity which  for
circular orbits is constant.~~$t_{diss}=(\omega-\Omega)R_1^3/\alpha G m_1$ is the dissipation time, 
with $\alpha$ the lag angle and $G$ the universal constant of gravitation.  Zahn (2008) suggests an approximation
$t_{diss} \approx R_1^2/<\nu>$, where $<\nu>$ is an average kinematical viscosity,  and we   
rewrite Eq. (\ref{tZsin}) as:  

\begin{equation}
\tau_{Zahn} = \frac{5\times 10^{-34}}{<\nu>} I \left ( \frac{\beta_0}{\beta_0-1}\right ) \frac{q^{-2}}{m_1/M_\odot}
\left (\frac{R_1}{R_\odot} \right)^{-6} \left (\frac{a}{R_\odot} \right)^{6}
\label{tZsin2}
\end{equation}

\noindent where $\tau_{Zahn}$ is in seconds.

In order to compare our synchronization times obtained from Eq. (\ref{tsin}) with those predicted in Eq. (\ref{tZsin2}), 
we now make the simplifying assumption of uniform density, from which $I = 2m_1 R_1^2/5$.


\begin{figure}
\centering
\includegraphics[width=0.80\linewidth]{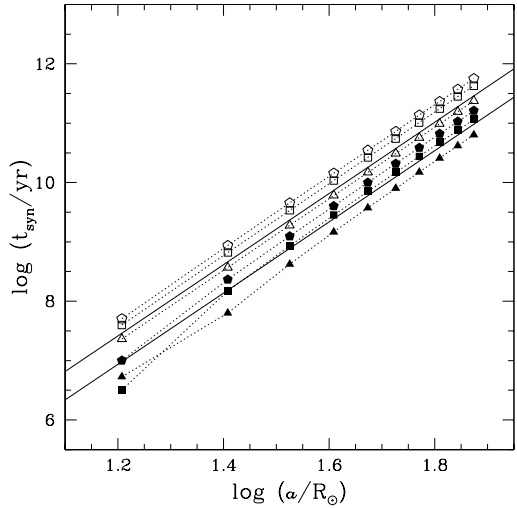}
\caption {Synchronization time computed with Eq. (\ref{tsin}) using
the results given in Fig. \ref{15874fg3}. The meaning of symbols stays
the same as in Fig. \ref{15874fg3}. The continuous lines are the
theoretical result obtained by using Eq. (2.5) from Zahn (2008), under
the assumption of a uniform density structure ($I=2m_1R_1^2/5$), and with
$<\nu>$=3.4$\times$10$^{12}$ cm$^2$ s$^{-1}$. The top  line is for
$\beta_0$=1.2  and the bottom one for $\beta$=2.0.
}
\label{15874fg7}         
\end{figure}

In Figure \ref{15874fg7} we show the values of ${\tau}_{syn}$ obtained from our
$\dot{E}$ values and Eq. (\ref{tsin}).  These results have approximately the 
predicted $a^6$ dependence, as obtained in  Toledano et al. (2007) with our 
earlier model.  We also see that the derived synchronization times 
depend on stellar rotation, with the faster rotating stars
($\beta_0$=2) attaining synchronization significantly faster than slower rotators 
($\beta$=1.2). 

 The two continuous lines are Zahn's theoretical result given by Eq. (\ref{tZsin2}) for 
the two values of $\beta_0$, using $m_1$ = 5 $M_{\odot}$, $m_2$ = 4 $M_{\odot}$, $R_1=3.2 R_\odot$,
and $<\nu>$=3.4$\times$10$^{12}$ cm$^2$ s$^{-1}$. Our results are  consistent with 
this theoretical result, except  that the value we used  for the kinematical 
viscosity in the calculations is significantly higher, 
$\nu$ =0.003 R$^2_\odot$ d$^{-1}$=1.7$\times$10$^{14}$ cm$^2$ s$^{-1}$.
It is important to note that in  our calculation the tidal friction comes  from the shear 
at the interface of each surface  element with its neighbors and with the inner 
(rigidly rotating) boundary.  This means that the action of the kinematical viscosity is 
limited to these shearing surfaces.   In addition, because ours is a single-layer approximation, 
there is no buoyancy, and therefore our viscosity values are not associated  with convection,
as are those of Zahn (1977, 2008).  Hence, at this stage, we are only able to draw an  
equivalence between the viscosity parameter used in our model and that used in Zahn's
formalism through the comparison of the synchronization timescales, as shown in Fig. \ref{15874fg7}.
That $<\nu>\sim$ 10$^{-2} \nu$ is likely because   in our 
calculations  the whole energy dissipation is forced to occur  in the surface layer, whereas
in Zahn's formalism deeper layers are also  involved in these processes. 



\subsection{Eccentric orbits}

\begin{figure}
\centering
\includegraphics[width=0.80\linewidth]{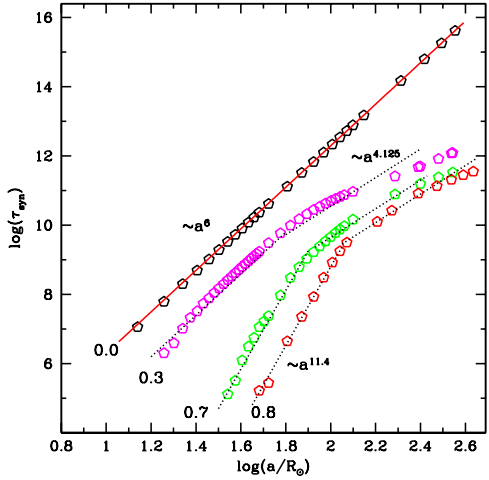}
\caption{Synchronization from the energy dissipation rates computed with our model using a
kinematical viscosity $\nu$=0.005 R$^2_\odot$ d$^{-1}$ and Eq. (\ref{tsinecc}). The
different ``families" are labeled by their corresponding eccentricity.  The solid line is the
synchronization timescale from Zahn's (2008) relation, assuming a constant density structure
and $<\nu>$=1.4$\times$10$^{12}$ cm$^2$ s$^{-1}$. This relation is valid only for $e\sim$0.0.
The dotted lines illustrate the trends for higher eccentricities from our computed models.
Note that $\tau_{syn} \sim a^{4.125}$ for high value of $e$ and $a$.  $\tau_{syn}$ is
in units of years.
}
\label{15874fg8}        
\end{figure}

Eq. (\ref{tsin}) can be conveniently written\footnote{Using $\Omega_0=1.428 \times 10^{-4} \vv_{rot}/R_1\beta_0$, 
and  $\vv_{rot}$=$\beta_0 R_1(1+e)^{1/2}/[0.02 P (1-e)^{3/2}$], where $\vv_{rot}$ is in km s$^{-1}$, P is in days, and
$R_1$ in R$_\odot$.} for an eccentric orbit in terms of the synchronicity parameter at periastron, $\beta_0$ and
the orbital eccentricity  $e$, as 
\begin{equation}
\tau_{syn} (yr)=3.17\times 10^{38} \frac{m_1R_1^2(\beta_0^2-1)(1+e)}{\dot{E}_{ave} P(1-e)^3},
\label{tsinecc}
\end{equation}

\noindent where $m_1$ and $R_1$ are the primary star's mass and radius, respectively, in solar units,
and $P$ is the orbital period in days.  The synchronization times obtained with the above
equation, using  the $\dot{E}_{ave}$ values
that are  plotted in Figure \ref{15874fg4} are shown in  Figure \ref{15874fg8}.  Several features of
this plot are notable: 1) there is a clear difference between moderately eccentric ($e\lesssim$0.3)
and the very eccentric systems; 2) at large orbital separations ($a\geq$100 R$_\odot$), the slope 
significanlty flattens  from the $\tau_{syn}\sim (a/R_\odot)^6$  that applies for circular orbits\footnote{
Note that for $a\geq$100 R$_\odot$, $\tau_{syn} \sim (a/R_\odot)^{4.125}$, similar to the synchronization
timescale predicted by Tassoul (1987).}; 
and 3) for a fixed value of $a$, the timescale for synchronization of the eccentric binaries is 
shorter than the corresponding circular orbit binary, with differences as large as five 
orders of magnitude visible in Figure \ref{15874fg8}.

\begin{figure}
\centering
\includegraphics[width=0.80\linewidth]{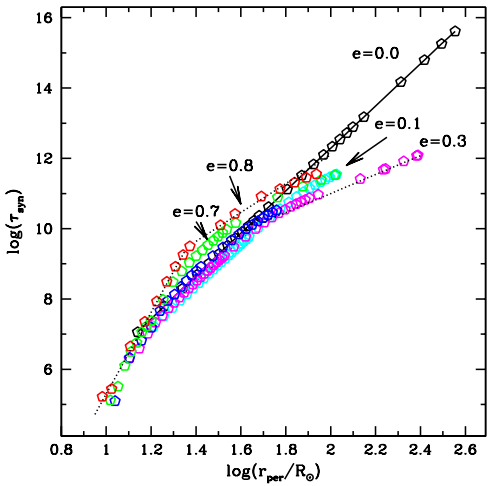}
\caption{Same synchronization times in the previous figure, but here
plotted as  a function of periastron distance, $r_{per}$.  The dotted lines
connect the points corresponding to each of the eccentricities.  $\tau_{syn}$ is
in units of years.
}
\label{15874fg9}         
\end{figure}

The separation into these different ``families" is eliminated in part by using the periastron distance, 
$r_{per}$, instead of the semi-major axis of the orbit. 
Figure \ref{15874fg9}  illustrates the same values of $\dot{E}$ shown in Figure \ref{15874fg8}, but in this 
case $\tau_{syn}$ is plotted as a function of $r_{per}$.  The different curves ``collapse" onto a 
common region in the $log(\tau_{syn}$) {\it vs.} $log(r_{per})$ plane, with significant differences occuring 
only for very large orbital separations.  This is consistent with the fact that, although the
mean distance between the two stars increases with $e$, the distance at periastron strongly
decreases, and as emphasized by Leconte et al. (2010),  most of the work caused by tidal forces 
occurs at this point of the orbit. 

The actual values of $\tau_{syn}$ depend on 
the  values of $\nu$ and  $k$. But  the scaling of $\tau_{syn}$ with separation 
at periastron is as plotted in Figure \ref{15874fg8}, unless $\nu$ and $k$ depend on orbital 
separation.\footnote{Because $\nu$ is generally associated with turbulence, and because 
the perturbations of the surface velocity field in short-period binaries is very strong, 
the turbulent component in these systems may be much stronger than in wider binary systems.  Thus, 
the use of a constant viscosity for all orbital separations may not be the most appropriate.}


\section{Orbital phase-dependence and periastron events}

Consider the binary system with $e$=0.8 and $P$=15d from the grid of models described in the
previous sections.  With $\beta_0$=1.2, the primary star, $m_1$, is rotating at 193 km s$^{-1}$.  
The maps in the top panels of Figure 10 are a color-coded Mollweide projection of the stellar surface, 
with white
corresponding to the maximum elevation from the unperturbed stellar surface and black to the largest
depression below the unperturbed stellar surface. Azimuth angle $\varphi$=0$^\circ$, corresponding to the
location of $m_2$, is on the left edge of the map, $\varphi$=360$^\circ$ on the right edge, and 
$\varphi$=180$^\circ$ is at the center.  
 At periastron (Figure 10, left), the surface acquires a shape that is typical of that assumed in 
the ``equilibrium tide" approximation. That is, a tidal bulge is raised on both hemispheres with the larger
bulge pointing in the general direction of the companion, $m_2$, but ``lagging" slightly behind.  The
maximum height  of the primary bulge at the equator (Figure 10, bottom left) is $\sim$0.03 R$_\odot$.\footnote{A rough 
estimate (c.f., Zahn 2008) of the tidal bulge elevation for a non-rotating star with similar characteristics, 
$\delta R/R_1 \sim (m_1/m_2)(R_1/r_{per})^4\sim0.02$, thus predicting $\delta R$=0.07 R$_\odot$,
a factor 2.5 larger than predicted by our model.} 


\begin{figure}
\centering
\includegraphics[width=0.40\linewidth]{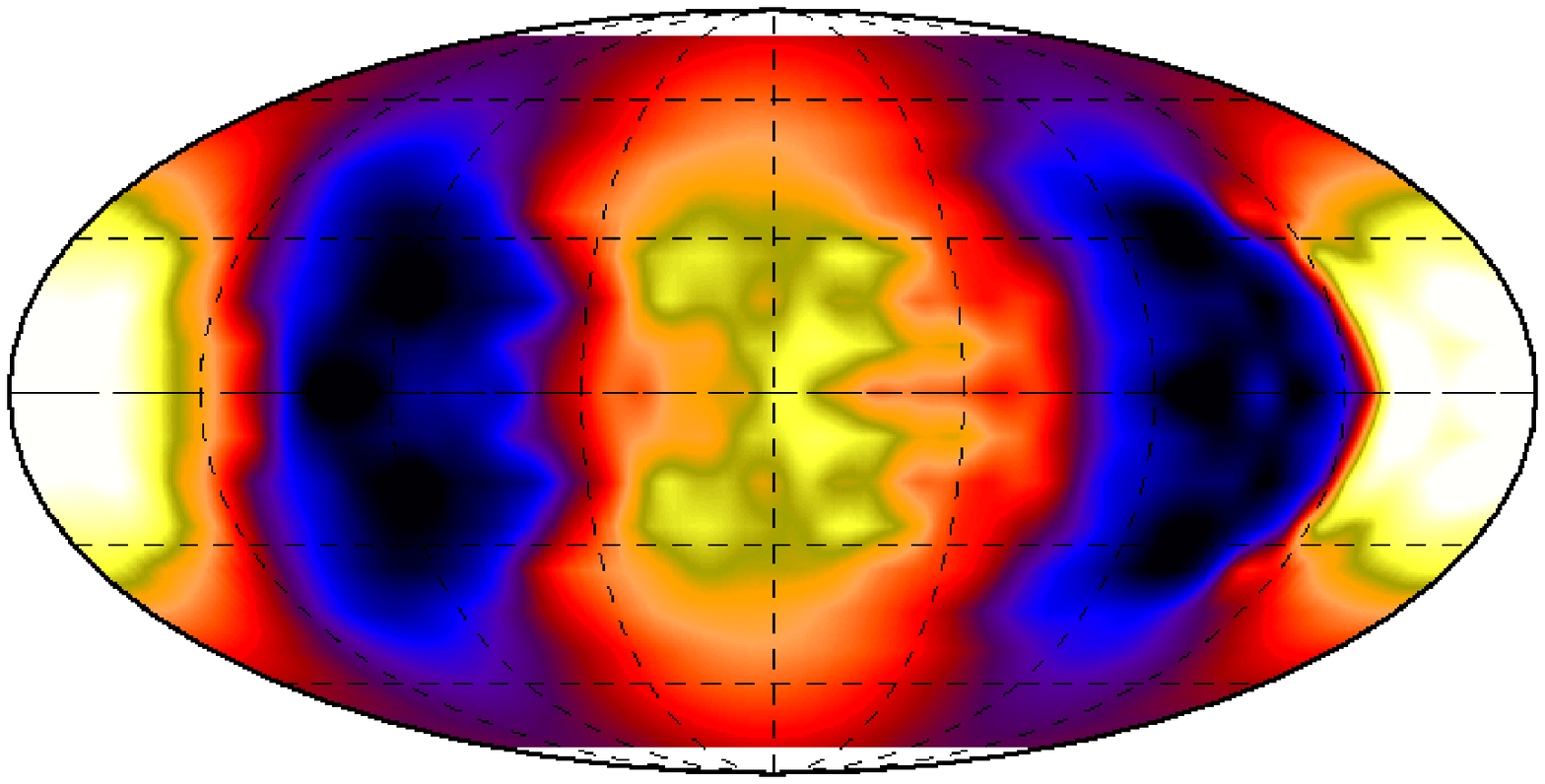}
\includegraphics[width=0.40\linewidth]{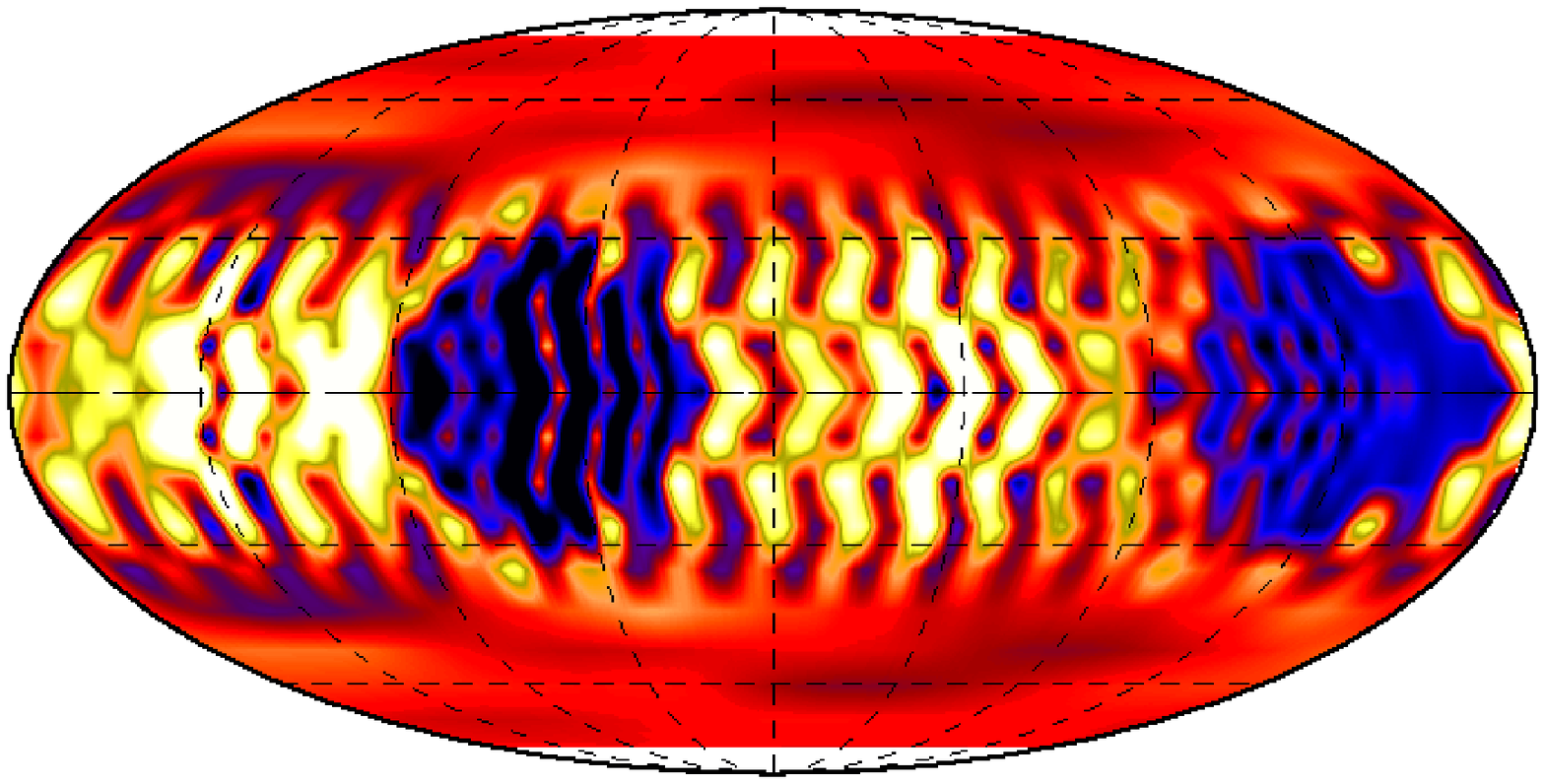}
\includegraphics[width=0.40\linewidth]{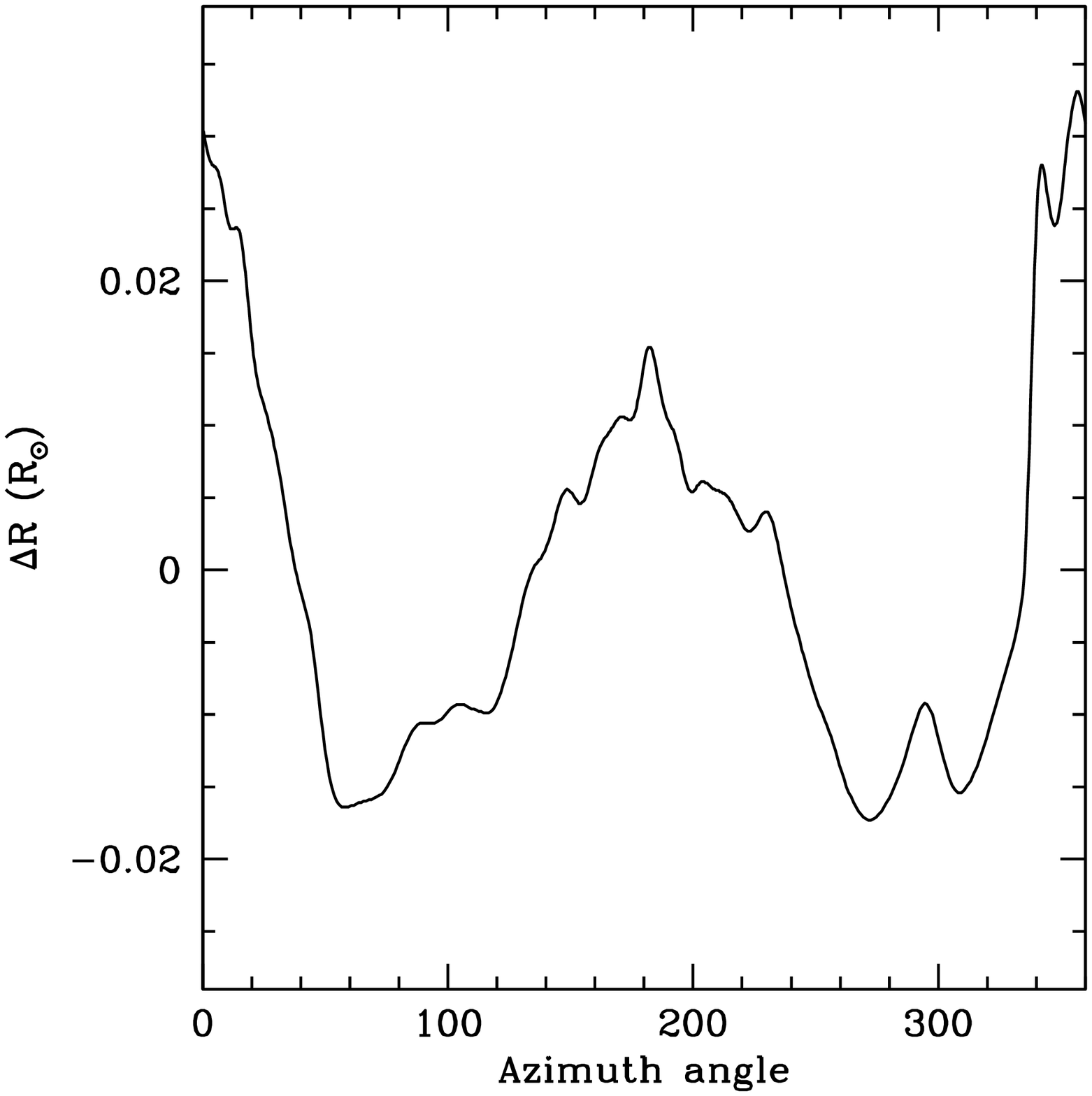}
\includegraphics[width=0.40\linewidth]{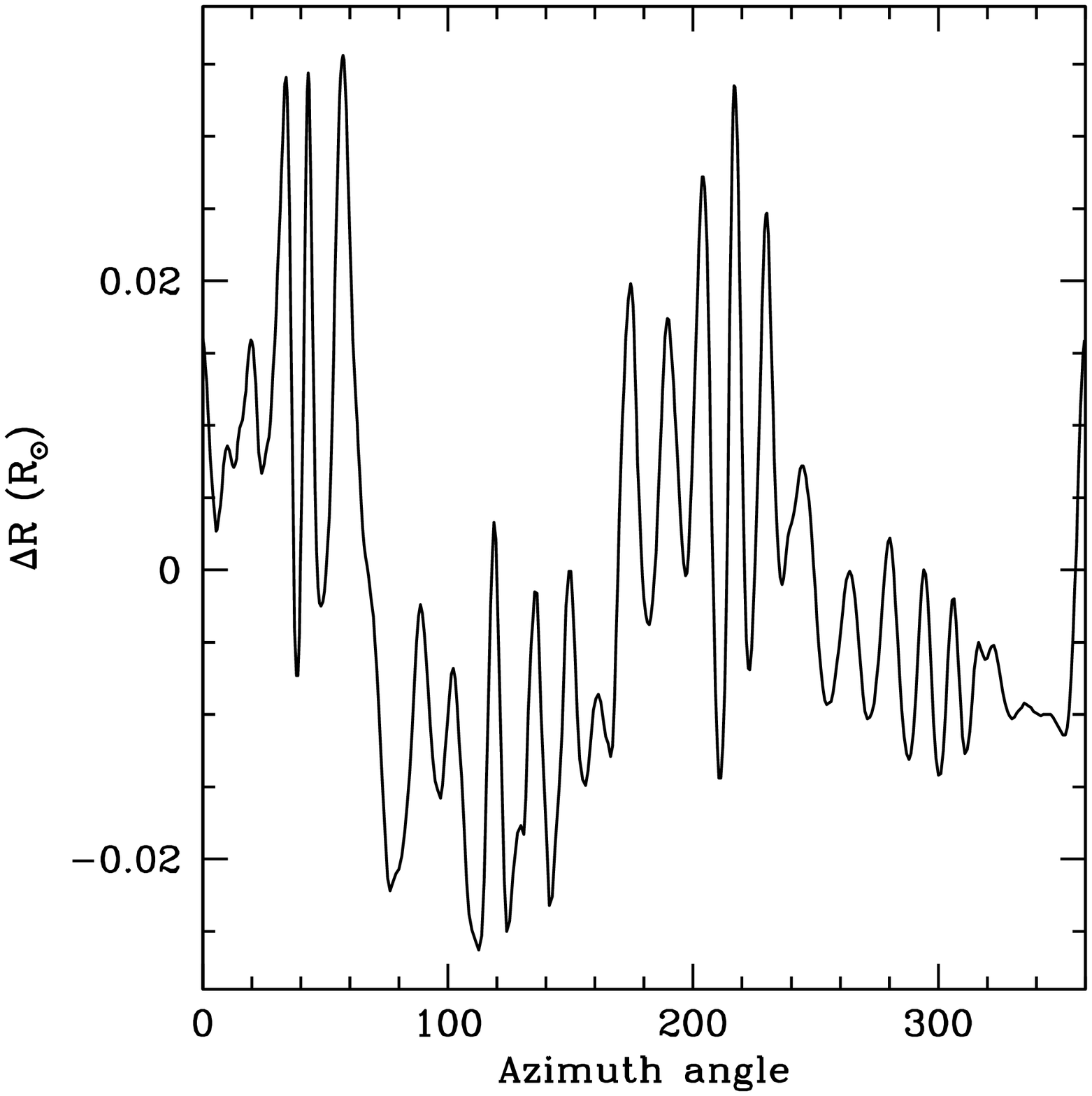}
\caption{Top: maps of radius residuals (r($\varphi$')$-R_1$) for e=0.8, P=15d, $\beta_0=$1.2 at periastron (left) and
0.833 days after (right). The sub-binary longitude ($\varphi$'=0) is at the far left on each map and
$\varphi$'=180$^\circ$ is at the center. The sense of the stellar rotation is from left-to-right.
White/black represents maximum/minimum residuals. Bottom: corresponding plots of the radius residuals along the
equator.
}
\label{15874fg10}          
\end{figure}

\begin{figure}
\centering
\includegraphics[width=0.40\linewidth]{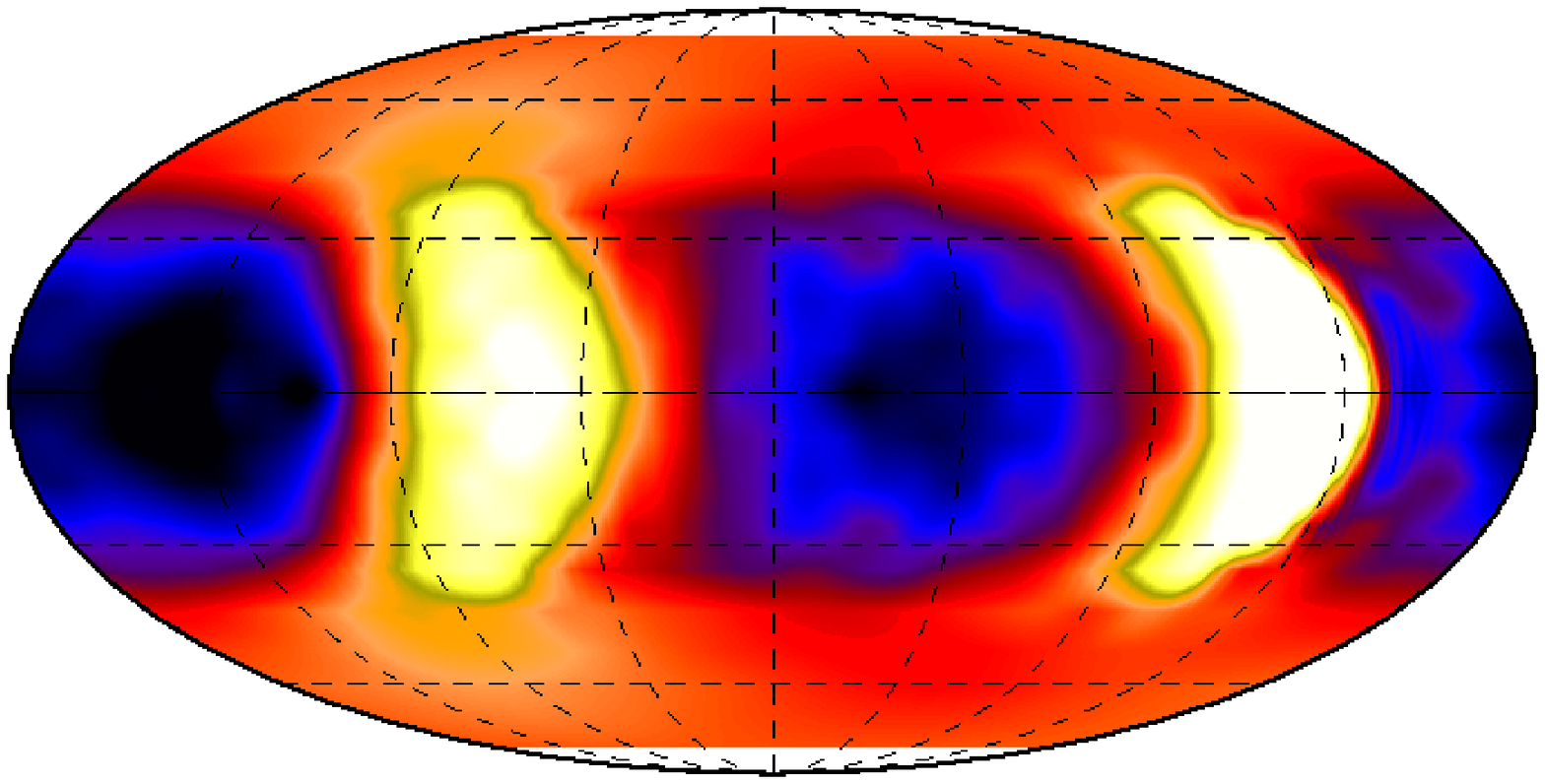}
\includegraphics[width=0.40\linewidth]{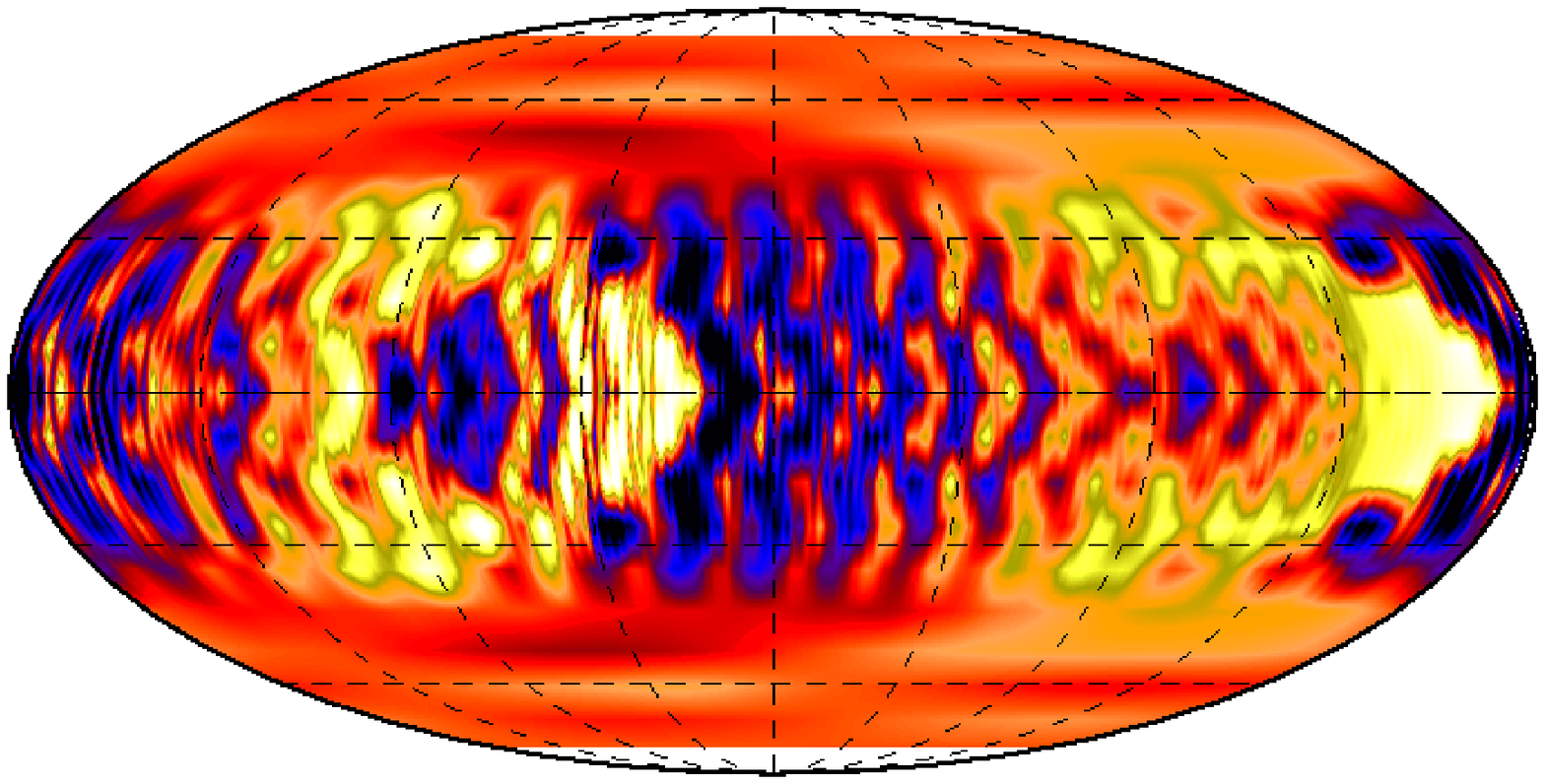}
\includegraphics[width=0.40\linewidth]{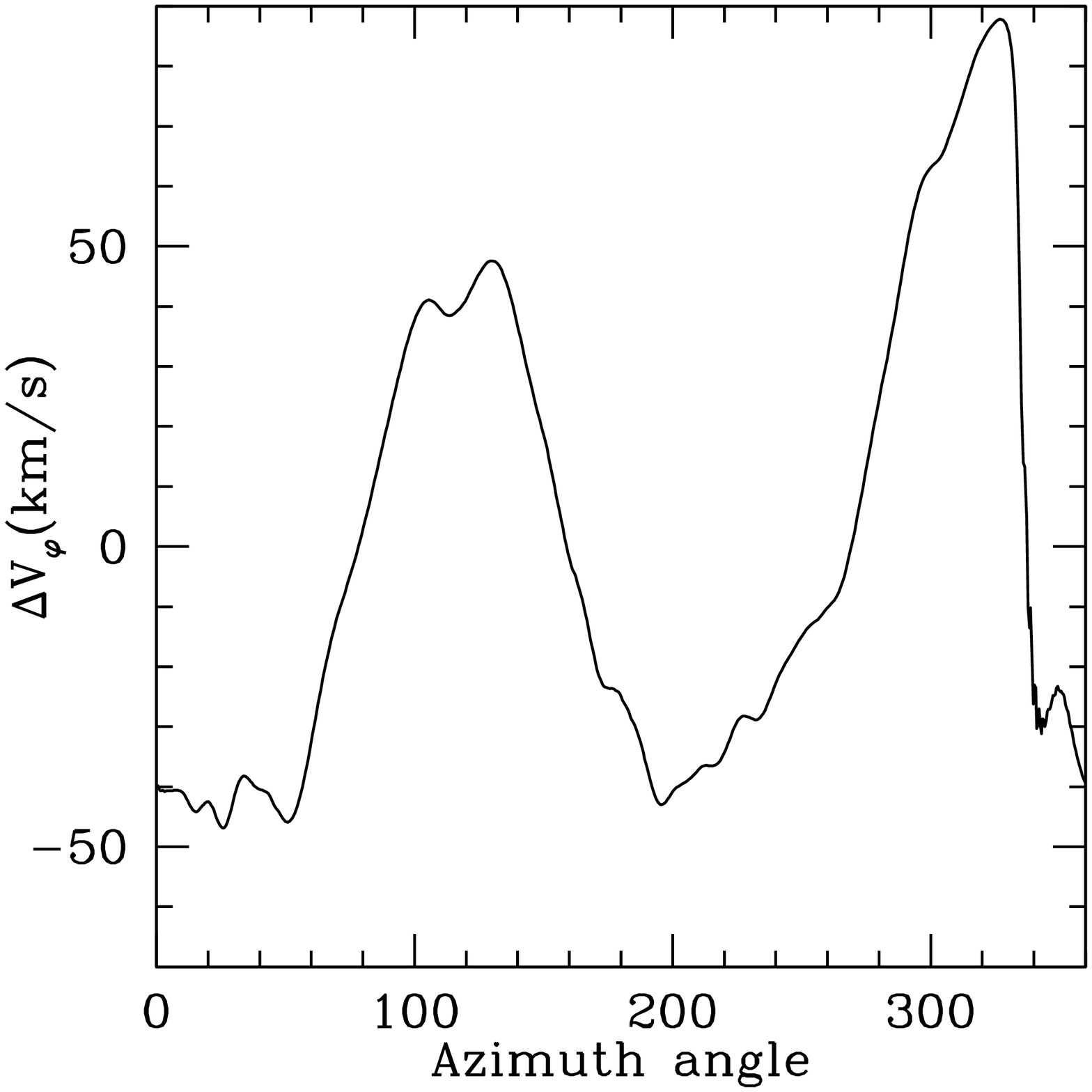}
\includegraphics[width=0.40\linewidth]{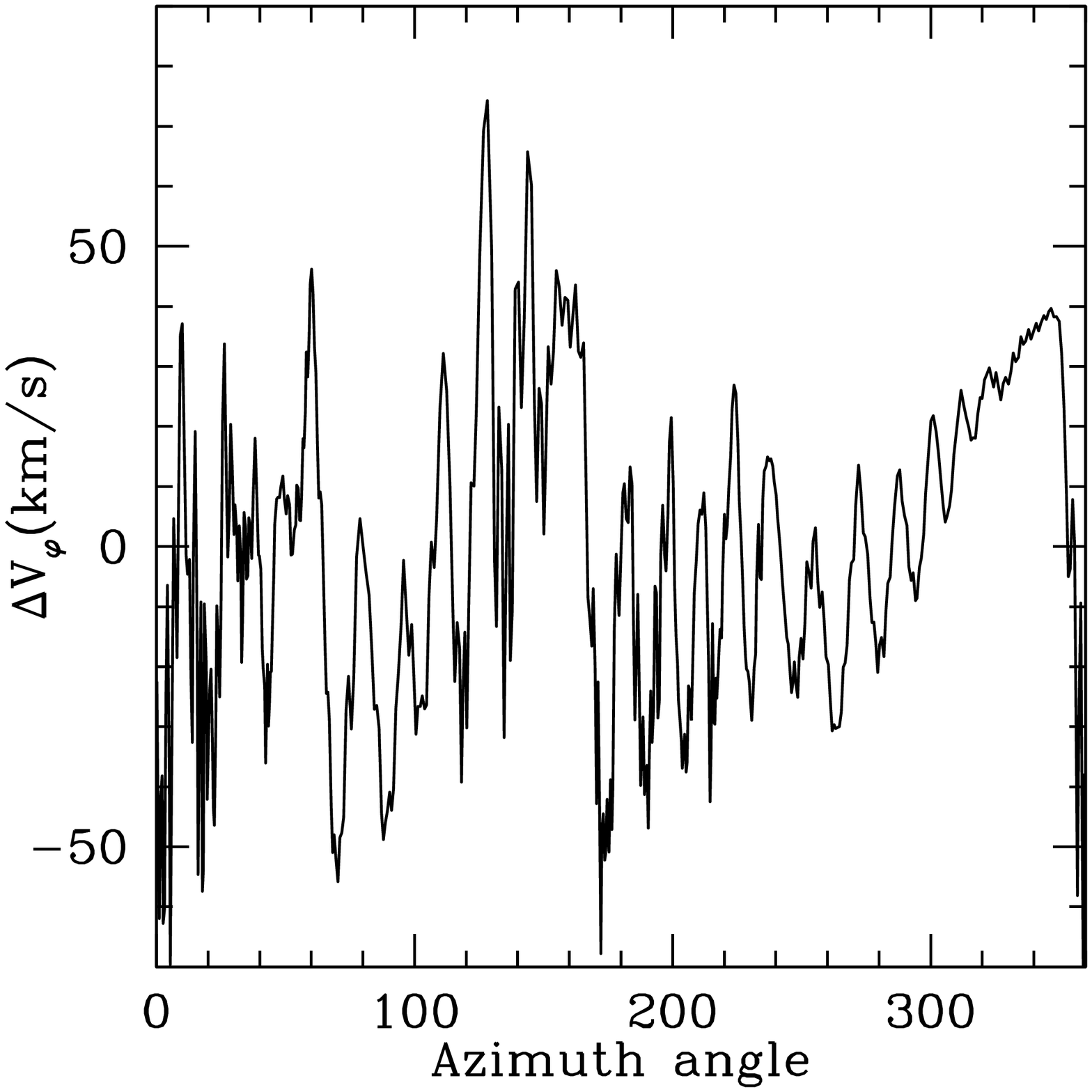}
\caption{Top: maps of tangential velocity residuals  for e=0.8, P=15d, $\beta_0=$1.2 at periastron (left) and
0.833 days after (right). The sub-binary longitude ($\varphi$'=0) is at the far left on each map and
$\varphi$'=180$^\circ$ is at the center. The sense of the stellar rotation is from left-to-right.
White/black represents positive/negative residuals; i.e., particles moving faster/slower than the
unperturbed stellar rotation velocity. Bottom: corresponding plots  along the
equator.
}
\label{15874fg11}     
\end{figure}

The ``equilibrium tide" configuration does not subsist very long, however.  The right-hand panel of Figure
\ref{15874fg10} illustrates the deformations just 0.833 day after periastron, and clearly  the
``equilibrium tide" approximation no longer represents the stellar surface shape.  Now it is
best described in terms of a large number of smaller-scale spatial structures.  These would fall 
into the ``dynamical tide" representation, if they were associated with
the normal modes of oscillation of the stellar interior.\footnote{Recall that the stellar interior here is 
a rotating, rigid body.} The amplitude of these structures declines with orbital phase
as apastron is approached.  A more in-depth analysis of the smaller spatial structures and their
dependence on the model input parameters  goes beyond the scope of the current paper, but the
important point to note is that  the ``equilibrium tide" approximation is valid only around
periastron passage for this particular highly eccentric model binary case.  We find that for systems in our
computed grid with low eccentricities and/or large orbital periods,  the  approximation 
remains valid over the orbital cycle, but not in the more extreme cases.  However, even 
there, the tidal bulges raised by the secondary cannot be assumed to be steady over the orbital
cycle, neither in amplitude nor in location.

We illustrate the azimuthal velocity field, $\vv_{\varphi'}$, in Figure  \ref{15874fg11}. The  
explanation given by Tassoul (1987) in the context of circular orbits is most appropriate to 
describe the phenomenon displayed at periastron (left): ``Assume that at some initial instant the 
primary is rotating rigidly with a constant supersynchronous angular velocity.  If there were no 
companion, the system would remain forever axisymmetric.  Because of the presence of the secondary, 
however, a fluid particle moving on the free surface of the primary will experience small 
accelerations and decelerations in the azimuthal direction.  Hence, in the sectors 
0$^\circ<\varphi'<\pi/2$ and $\pi<\varphi'<3\pi/2$, 
the azimuthal component $\vv_\varphi'$,  of this fluid particle will be smaller than the typical 
value by a small amount. On the contrary, in the other sectors, $\vv_\varphi'$ will be larger."  Note 
that in the case shown in Figure \ref{15874fg11} the pattern of positive and negative residuals is
slightly rotated with respect to Tassoul's description because this is not a circular-orbit case, and
also because of the ``lag" of the primary tidal bulge.  This quadrupolar pattern rapidly breaks down
into smaller scale structures after periastron, corresponding to the smaller spatial structures of 
Figure \ref{15874fg10} (right).

\begin{figure}
\centering
\includegraphics[width=0.40\linewidth]{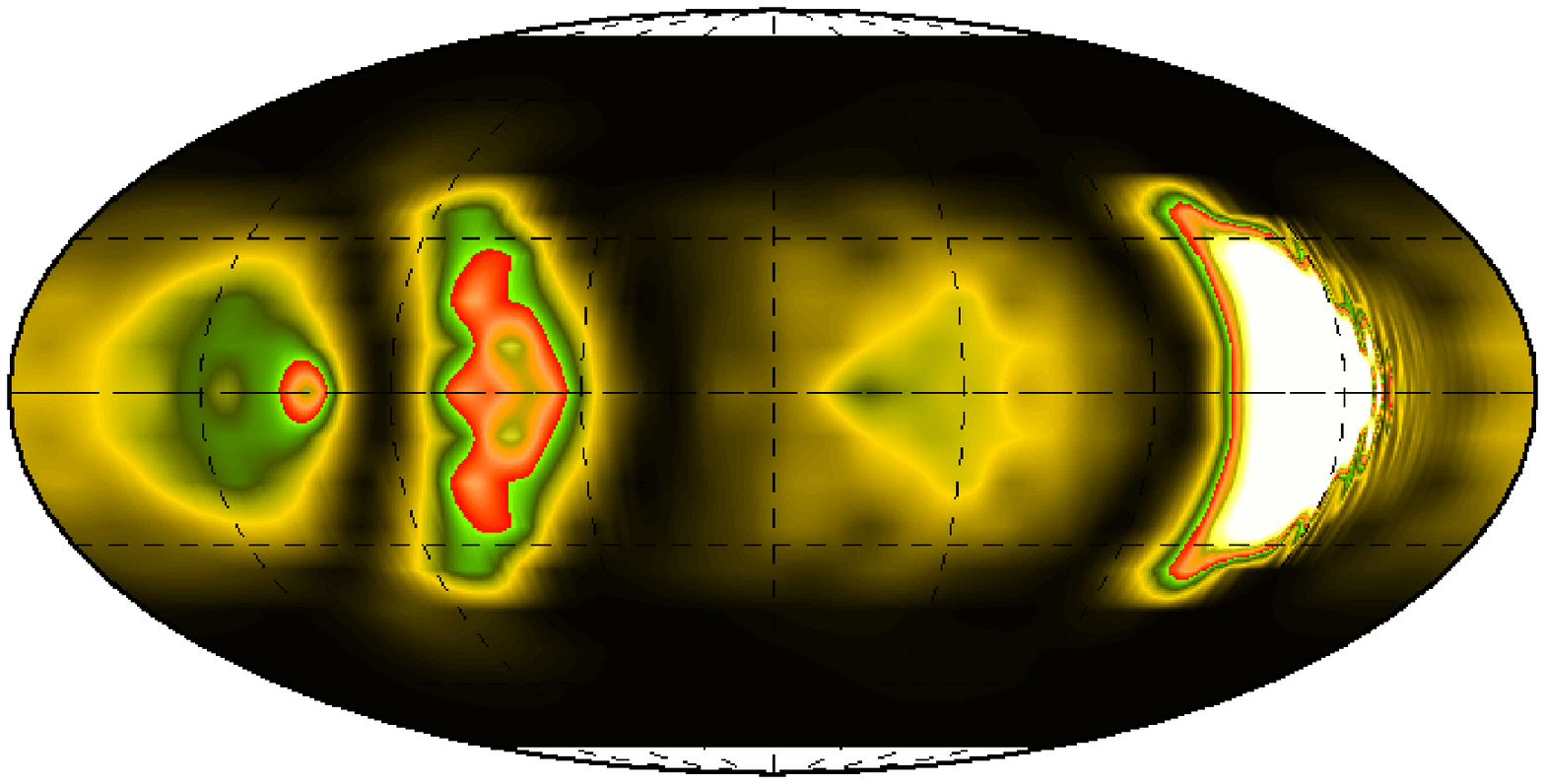}  %
\includegraphics[width=0.40\linewidth]{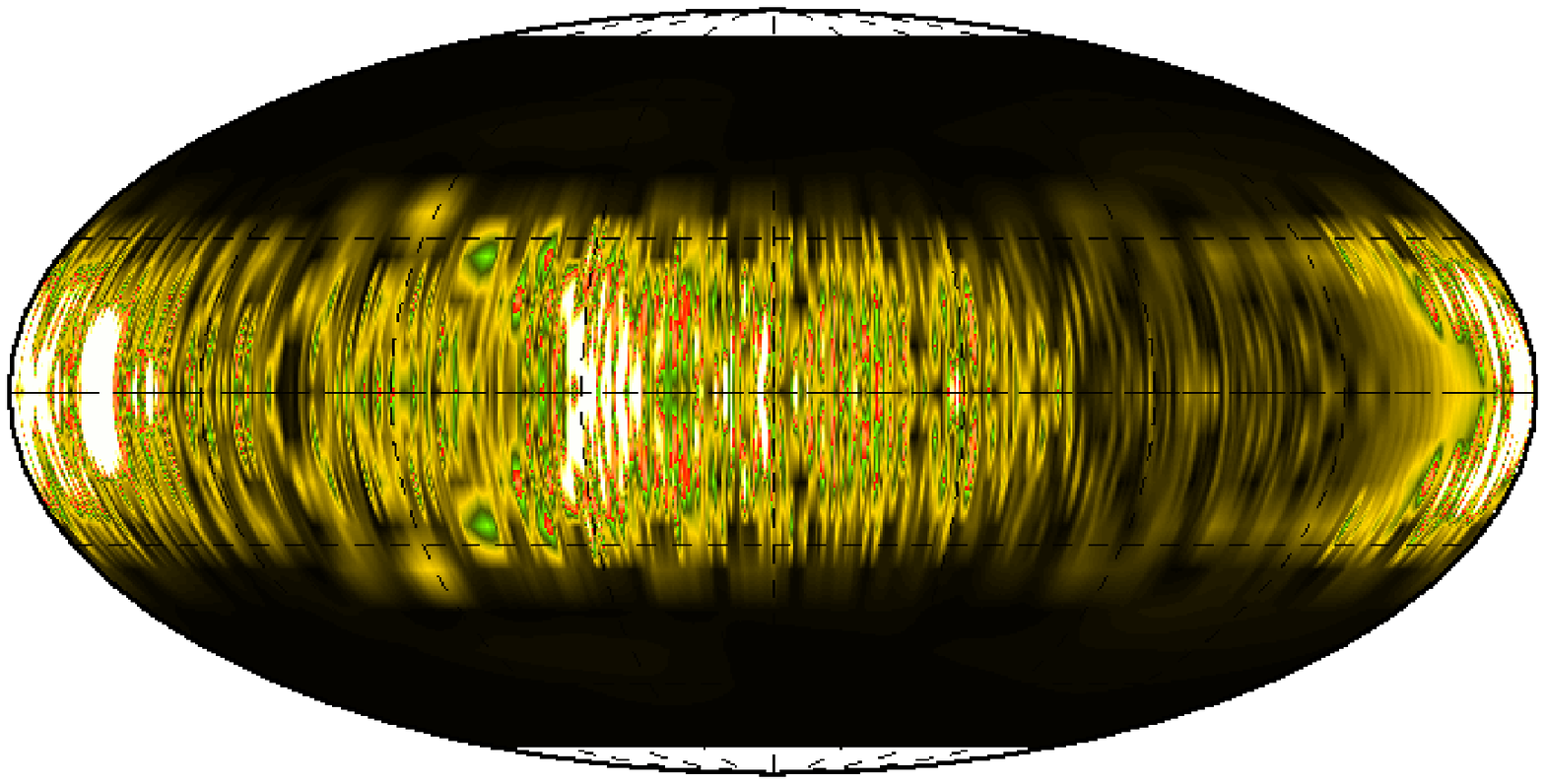}
\includegraphics[width=0.40\linewidth]{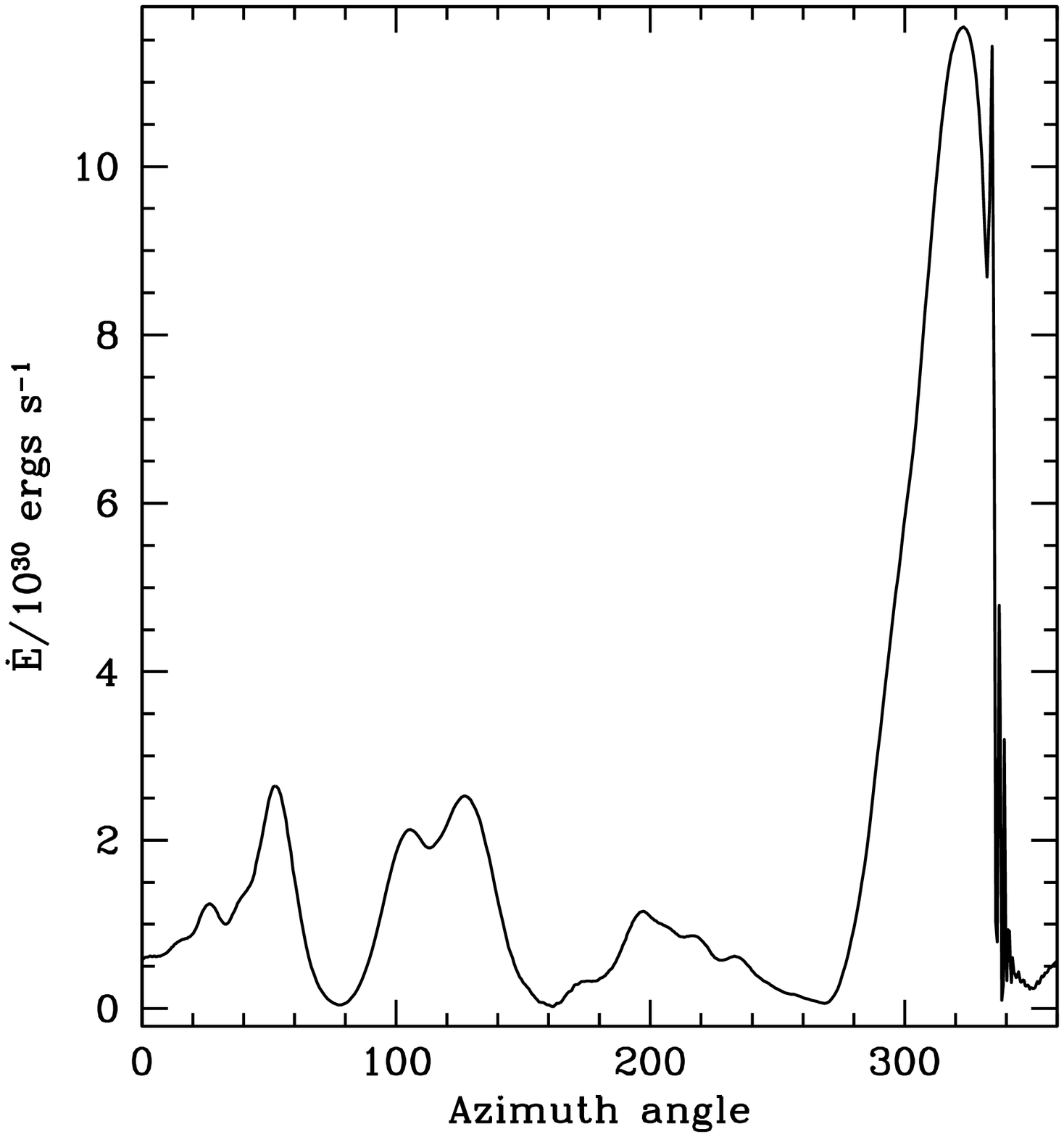}
\includegraphics[width=0.40\linewidth]{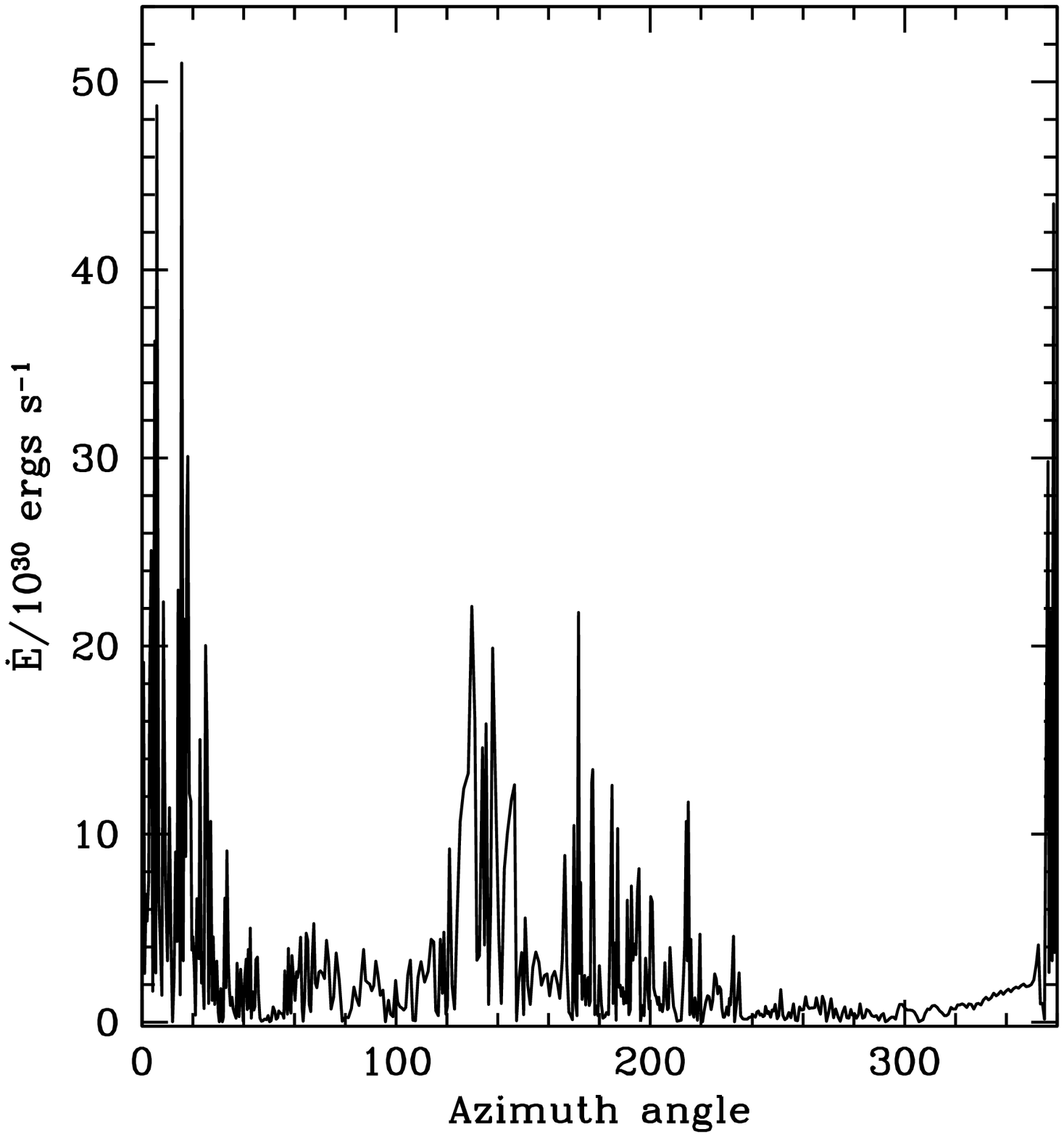}
\caption{Top: maps of energy dissipation rate  for e=0.8, P=15d, $\beta_0=$1.2 at periastron (left) and
0.833 days after (right). The sub-binary longitude ($\varphi$'=0) is at the far left on each map and
$\varphi$'=180$^\circ$ is at the center. The sense of the stellar rotation is from left-to-right.
White/black represents maximum/minimum residuals. Bottom: corresponding plots  along the
equator.
}
\label{15874fg12}            
\end{figure}

Highest energy dissipation rates are associated with the regions having large perturbations in
the  azimuthal motion, both positive and negative. At periastron (Fig. \ref{15874fg12}, left), 
the highest $\dot{E}$ values occur close to the sub-binary longitude. But by 0.833 day later
(Fig. \ref{15874fg12}, right), the large regions where most
of the $\dot{E}$ is generated have broken down into small spatial scale structures.  It is
noteworthy, however, that some of these smaller structures are associated with significantly
high values of $\dot{E}$.   Also noteworthy is that nearly the entire $\dot{E}$ occurs
at $\pm$45$^\circ$ from the equatorial latitude.

\begin{figure}
\centering
\includegraphics[width=0.80\linewidth]{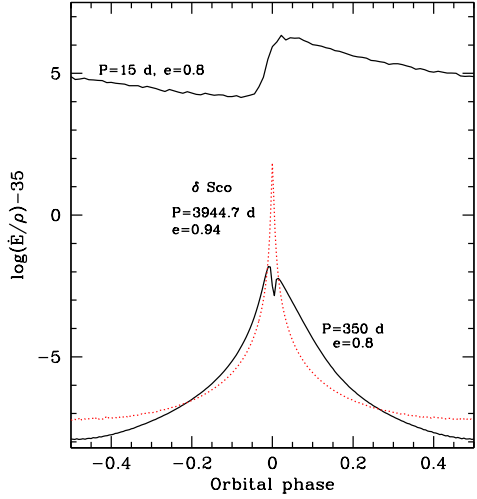}
\caption{ Tidal shear energy dissipation rate for  very eccentric binary systems as a
function of orbital phase measured from periastron passage.
$\dot{E}$ is in ergs s$^{-1}$ and $\rho$ in gm cm$^{-3}$.  Shown in this
figure are two cases with e=0.8 (P=15 and 350 days) and a case for $\delta$ Sco.
Note that peristron passage is always associated with a very large time
variation of $\dot{E}$.
}
\label{15874fg13}        
\end{figure}

The main conclusions we draw from this example are that 1) periastron passage causes a sudden and brief
change in the surface perturbations, leading to significant dynamical effects; 2) the tidal shear
energy dissipation rates are concentrated (in this example) around the equatorial latitude.  Hence,
the question arises as to whether these phenomena may provide a natural explanation for the
periastron events described in the introduction, as well as the creation of a disk from the 
ejected material.  The topic of how the material may be ejected goes beyond the scope of this
paper\footnote{See, for example, Lee (1993)}, but a plausible scenario is that additional energy 
is fed into sub-photospheric layers by the tidal shear, combined with the possible additional turbulence 
caused by the rapidly-changing  motions on the stellar surface.
We suggest that one of the crucial ingredients is the extremely rapid and strong forcing to which 
the stellar surface is subjected during periastron passage in very eccentric systems.  Figure 
\ref{15874fg13} shows 
a plot of $\dot{E}$ as a function of orbital phase for the $e$=0.8, $P$=15 d binary model discussed 
above and two other cases for comparison.\footnote{Note that actual binary systems with this eccentricity and
orbital period are extremely rare or non-existent, as can be seen in Fig. 1 of Mazeh (2008), which
is probably because of the very large $\dot{E}$, which should rapidly tend to circularize the orbit.}
The rapid rise in the $\dot{E}$ value at periastron is evident.

The second case depicted in Figure \ref{15874fg13} corresponds to a $e$=0.8, P=350 d model from our grid, and
the unperturbed equatorial rotation velocity of $m_1$ is  $\sim$8 km s$^{-1}$.  The $\dot{E}$
values undergo a $\sim$6 order-of-magnitude variation between apastron and periastron, most of
which occurs within $\pm$70 days of periastron passage.  Finally, the third case illustrated in
Figure \ref{15874fg13} is an exploratory model calculation for the extremely eccentric ($e$=0.94) 
Be-type binary system $\delta$ Sco.\footnote{The parameters used for this calculation 
are  $m_1$= 15 M$_\odot$, $m_2$= 2 M$_\odot$, $P$=3944.7 d, $R_1$= 7 R$_\odot$, $\beta_0$=28.5, 
$\Delta  R_1/R_1$=0.06, $\beta_0$=28.5, $\nu$=0.08 R$^2_\odot$ d$^{-1}$} 
The primary star in this system has an estimated equatorial rotation velocity of $\sim$200 km s$^{-1}$,
which means that  it is rotating extremely fast, compared to the orbital angular velocity of $m_2$. 
The star undergoes an increase in $\dot{E}$ by $\sim$9 orders of magnitude between apastron and
periastron, with a very steep rise to maximum just 15 days before periastron.


\section{Discussion}\label{disc}

A large body of research, starting with Darwin (1880), exists on the
effects produced by tidal interactions on the long-term evolution of
binary systems. The primary concern of these investigations is generally the
timescales for circularization of the orbit and synchronization of the
rotational and orbital angular velocities.  The three general approaches 
that have been developed for analyzing the effect of tidal dissipation in
these investigations are  1) the ``equilibrium tide" model; 2) the
``dynamical tide" model; and 3) the hydrodynamical model of Tassoul (1987). 

In the ``equilibrium tide" model (Darwin 1880; Alexander 1973; Hut 1980, 1981;
and see Eggleton, Kiseleva \& Hut 1998 and references therein)\footnote{An illustrative 
description and discussion of the limitations of this model may be 
found in Ferraz-Mello et al. (2008).} the shape of the stellar surface is 
approximated to that which it would adopt in equilibrium
under the effect of the  gravitational-centrifugal potential of the
system. The viscous nature of the stellar material leads to a misalignment (refereed to as ``lag")
between the  axis that connects the two tidal bulges and the line that connects
the centers of the two stars.  The action of the external gravitational field on this
misaligned body leads to a  torque which, in turn, leads to an interchange of energy and 
angular momentum between stellar rotation and orbital motion. Dissipation by viscosity  
results in  the decrease in the eccentricity of the orbit and  the synchronization
of the rotational and orbital motions.  It is important to note that the
``equilibrium tide" approach implicitly assumes that the star 
goes through a succession of rigidly rotating states, until the equilibrium
configuration is reached.

The ``dynamical tide" approach is based on the assumption that the star
may be viewed as an oscillator whose normal modes may be excited by
the variation in the gravitational potential.  Among the first mathematical 
formulations available in the published literature are those of Fabian, 
Pringle \& Rees (1975), Zahn (1975), and Press \& Teukolsky (1977).  In more 
recent years, a large amount of effort has been invested in further developing 
these concepts, as reviewed by  Savonije (2008)  (see, also McMillan, McDermott \& 
Taam 1987; Goldreich \& Nicholson 1989; Dolginov \& Smel'Chakova 1992; Lai 1996; 
Mardling 1995; Kumar \& Goodman 1996; Kumar \& Quataert 1997).
Three types of modes are believed to be excited a) the internal gravity modes, where the
restoring force is the bouyancy force in stably stratified regions
(Zahn 1975; Zahn, 1977; Goldreich \& Nicholson 1989; Rocca 1989);            
b) the acoustic modes, where the restoring force is the compressibility of
the gas; and c)  the inertial modes where the restoring force is the
Coriolis force (Witte \& Savonije, 1999a, 1999b). These modes are the
mechanism by which energy from the orbital motion is absorbed by the star. The 
system is driven toward the equilibrium state by the damping of these modes.   
Hence, it is necessary to make an assumption concerning the damping
mechanism. In stars with an outer convective layer, it is  generally 
assumed that turbulent viscosity is the dominant mechanism.  In stars with
outer radiative layers, radiation damping is assumed to be the dissipative 
mechanism. It is interesting  that in an eccentric system, the gravitational
forcing  goes through a large range of frequencies as the orbit progresses from apastron to periastron.
Hence, there is a good chance that one of these frequencies will resonate with the star's normal
modes, thus exciting the particular oscillation mode in the star.  At the same time,
as periastron is approached, however, the star becomes more deformed, thus potentially modifying its
normal modes.  
The above methods rely on a mathematical formulation of the star's 
response to the perturbing potential of its companion, and are usually
treated in the linear regime. Thus, they are generally limited to slow
rotational velocities and/or low orbital eccentricites.

The third approach, introduced by Tassoul (1987) for early-type  binaries,
involves large-scale hydrodynamical motions within the tidally distorted star.  
It is based on the assumed presence of large-scale meriodional circulation, 
superposed on the mean azimuthal motion of stellar rotation. This circulation 
is believed to arise because a fluid particle moving on the free surface of the 
primary will experience small accelerations and decelerations in the azimuthal 
direction, depending on its location with regard  to the sub-binary longitude.\footnote{
Indeed, our simulations show that these azimuthal motions are the dominant cause of
photospheric line-profile variability on orbital timescales (Harrington et al. 2009).}
These meridional circulation currents  transport angular momentum very 
efficiently, which leads to very rapid synchronization timescales; i.e., 
the spin-down time primarily depends on (a/R)$^{4.125}$.  This mechanism is thus
a  more long-range mechanism compared with the other two.  The method recognizes that 
near the surface, the Coriolis force and the viscous force must play a donominant role 
in the mechanical balance. Hence, this approach significantly differs from the other 
two in that it captures the hydrodynamic processes near the stellar surface.

Contrary to the three methods described above,  where the primary concern is the long-term 
($\geq$10$^6$ yr) evolution of the orbital and stellar rotation parameters, our method
was developed to analyze the tidal effects on orbital timescales ($\sim$days).  However, we are
able to compare the results of our model with results from the other models through the energy 
dissipation rates which, in turn, allow an estimate to be made of the synchronization
timescales.  We find that the results of our model for circular orbits are fully consistent with 
the results of Zahn (1977, 2008) for  stars with convective envelopes.  We also find that       
for eccentric orbits, our results for the scaling of the energy dissipation rates with
orbital eccentricity and separation are consistent with the predictions of Hut (1981). 

Zahn's synchronization timescale for convective envelopes 
is based on the ``dynamical tide" formalism, while Hut's method is based on the
``equilibrium tide" approximation.  Our method  does not rely on any {\it a priori} assumption 
regarding the mathematical formulation of the tidal flow structure. Instead, we solve the
equations of motion of a thin, deformable surface layer that is
allowed to respond to the gravitational field of $m_1$ and $m_2$,
the Coriolis force, the centrifugal force, the gas pressure and the viscous forces
from the shearing flows. The numerical solution of this system of
equations yields the velocity field on the stellar surface, from which
the energy dissipation rates may be computed. 
Thus, ours is a computation from first principles, the only
simplifying assumptions being those which are inherent to the one-layer 
approximation. In particular, our method neglects the deformations
produced by the tidal effects on the layers below the surface layer.
However, because the tidal forces decline as 1/r$^{3}$, the tidal
effects are weaker in deeper layers of the star than on the surface
(see, for example, Dolginov \& Smelch'akova 1992).   On the other hand, we are fully
aware that the detailed behavior of underlying layers will be coupled to that of
the surface layer.  Thus, at this stage our results are merely indicative of the 
detailed behavior of the surface layer as a function of the  orbital phase.

Among the advantages of our method is that it allows a qualitative assessment of the response 
of the stellar surface under arbitrary conditions and  it   
captures the essence of the three general approaches described above.  That is, the 
solution of the equations of motion  yields a first-order deformation attributable to the 
equilibrium tide and a lag angle; a second order deformation attributable to  dynamical effects 
\footnote{Clearly, the dynamical effects predicted by the model refer to the surface layer only.}; 
and because of the inclusion of gas pressure, it  captures some of the hydrodynamical effects.
Interestingly, our results for the  synchronization timescales in eccentric orbits are consistent
with the equilibrium tide approach at intermediate orbital separations, but suggest much faster
synchronization timescales for larger orbital separations, consistent with Tassoul's hydrodynamic 
model.\footnote{Note, however, that our model does not currently include meridional motions.  The connection
with the Tassoul model resides in the concept of the azimuthal flow velocities.  These have
significantly faster speeds   than those of meridional flows and  act on very short timescales.}
From the observational standpoint, Abt \& Boonyarak (2004)  found that B and A stars in binaries with periods 
as long as 500 days have rotational periods significantly shorter than the corresponding single stars.  
They therefore concluded that the synchronization process can have an inpact on the stellar rotation even for
relatively wide binaries with periods as long as 500 days, thus supporting the presence of a longer-range
mechanism.  Claret et al. (1995) and Claret \& Cunha (1997) also have noted that
the Zahn models and the Tassoul models each can explain the observational data only in part. 

The major limitation of a one-layer model lies in  neglecting the oscillatory nature of  the 
stellar interior.  In particular,  the interaction of the normal modes of oscillation of the star 
with the external forcing frequency provides a  means by which  orbital energy might
be captured by the star and deposited as thermal energy. Hence, we cannot at this time probe
interesting situations associated with  resonances.  On the other hand,  in
long-period, very eccentric binary systems the peristron passage involves very large temporal
gradients in  the surface properties of the star.  The tidal shear energy dissipation rates, in 
particular, change dramatically over a very small portion  of the orbital cycle.  Our
exploratory calculation for the $\delta$ Sco binary system (orbital period P=3944.7 days) 
indicates that $\dot{E}$ changes by five orders of magnitude over the 82 days before periastron,  
two of these occurring over the 15 days before peristron. The combination of added energy, extended
radius, and high surface flow velocities may be envisaged to produce a destabilizing effect that leads 
to a mass-ejection event.  Interestingly, we expect that the ejecta should  form a disk-like structure,
because the strongest tidal effects occur in the equatorial plane.  Observational evidence 
for such a disk-forming event actually exists for the previous orbital cycle (Miroshnichenko et 
al. 2003).  Confirming this behavior in the upcoming periastron passage of 2011 would  strengthen 
the hypothesis.  

Finally, we note that it may be productive to use the periastron distance, $r_{per}$, instead of 
the semi-major axis, $a$,  when comparing statistical studies of observational properties with
theoretical models of tidal evolution.
We show that by using $r_{per}$, the synchronization timescales for binaries with different
orbital eccentricities ``collapse" into a small region in the $\tau_{syn}$ {\it vs} $r_{per}$
diagram, whereas on the $\tau_{syn}$ {\it vs} $a$ diagram, there is a different ``family" of
curves corresponding to each eccentricity.  This means that for a fixed value of $a$, the
synchronization timescale derived from eccentric binary systems is significantly  shorter (by
orders of magnitude) than a corresponding $e$=0 system.  However, if $r_{per}$ is used instead,
the spread between systems with different eccentricites is smaller.

We emphasize moreover  that there is  a significant dependence of $\tau_{syn}$ on the stellar 
rotational velocity.  However, it must be kept in mind that the important parameter is not only 
the equatorial rotational velocity, but also the synchronicity parameter, $\beta_0$.  Hence, slowly 
rotating stars may undergo strong tidal interaction effects if, for example, $\beta_0<<1$.  Whether 
strong tidal effects can contribute to alter chemical abundances is a question that merits
examination, because this could potentially be an explanation for some of the discrepancies
recently found between stellar structure evolutionary models and observations (Hunter et al. 2008; Brott 
et al. 2010).  

Clearly,   many of the outstanding questions associated with tidal interactions require a more 
complete treatment than that of any of the currently developed models. In particular, the departures 
from spherical symmetry imply that a one-dimensional treatment of the radiative transfer processes is 
inadequate.  Furthermore,  the energy that is dissipated due to tidal interactions, regardless of the 
mechanism invoked, is an added source of energy that is deposited in sub-photospheric layers of the star 
and needs to be considered when analyzing  energy transport processes and determining the structure of 
the outer stellar layers. 

\section{Conclusions}

We derived a mathematical expression for computing the rate of dissipation, $\dot{E}$,
of kinetic energy by the viscous flows that are driven by tidal interactions  and used
it to explore the effects caused by periastron passage in  eccentric binary systems.
Our method does not rely on any {\it a priori} assumptions regarding the mathematical 
formulation for the  perturbations in the star caused by its companion.
We show that our  results for the scaling of $\dot{E}$ with  orbital separation for circular 
orbits are the same as those derived analytically by Hut (1981), and that our results yield
the same scaling of synchronization timescales with orbital separation as found by Zahn (1977, 2008)
for convective envelopes.  For elliptical orbits, our results reproduce  the dependence of
$\dot{E}$ on the eccentricity  and  the general trend with orbital separation as predicted
by Hut (1981).  In addition, the model also adequately predicts the pseudo-synchronization 
angular velocity  for moderate eccentricities ($e\leq$0.3), as obtained by Hut (1981).


We explored  the distribution of $\dot{E}$ over the stellar surface, and found that its
configuration in small eccentricity binaries is consistent with the quadrupolar morphology 
of the ``equilibrium tide".  That is,  maximum  $\dot{E}$ values tend to be  located at 
four longitude regions where the fastest azimuthal motions are present. 
In very eccentric systems, however, the morphology significantly changes  as a function of 
orbital phase. Although generally quadrupolar-like at periastron, shortly after this time  
$\dot{E}$ is distributed into smaller spatial structures, which suggests an analogy with the 
``dynamical tide" models.  In addition, the perturbations are in general strongest around 
the equatorial latitude.\footnote{The equator coincides with the orbital plane  in our models.}

We speculate that the  very large and sudden increase in $\dot{E}$ that occurs around periastron,
combined with the rapid growth rate of the velocity perturbations is a promising
mechanism for explaining the increased stellar activity and mass-ejection events often observed
around this orbital phase.  Two factors may be involved 
in causing the periastron events.  First,  $\dot{E}$ is an additional
source of energy that is deposited in sub-photospheric layers and represents a time-dependent
perturbation to the temperature and differential velocity structure of these layers.  

Second, in very eccentric binaries,  the growth rate  of the surface perturbations increases very abruptly
at periastron. For example, in the very eccentric binary system $\delta$ Sco ($e$=0.94, $P$=3944.7 d),  our
exploratory calculation indicates that $\dot{E}$ changes by five orders of magnitude over the 82
days before periastron, with a two-orders of magnitude increase over the 15 days before peristron. 
The qualitative phenomenon is one in which portions of the stellar surface are forced to suddenly
rotate faster or slower than their neighboring inner layers, thus leading to a variable 
differentially-rotating structure which significantly differs from its 
unperturbed structure. Furthermore, most of the activity occurs around the equatorial latitude.   
The consequence for the stability of the outer layers under these conditions is a problem that 
merits further attention.


\begin{acknowledgements}
We express our gratitude to Frederic Masset for a critical reading of this paper
and very helpful suggestions, and to the referee, Peter Eggleton, for
important comments.  We also thank J. Kuhn, P. Garaud, and A. Maeder for
motivating and helpful discussions;  I. Brott for  the 5 M$_\odot$ stellar model;
and  Alfredo D\'{\i}az and Ulises Amaya for computing assistance.
This research was funded under grants DGAPA/PAPIIT/UNAM 106708; CONACYT 48929; travel
support to GK and DMH from the CONACYT-NSF collaborative program  is acknowledged.
\end{acknowledgements}

{}

\begin{appendix}
\section{Grid of model calculations}
Table 1 lists the input parameters of the grid of model calculations used in this paper.
The first column identifies the set number; the second column lists the eccentricity; the
third column lists the range in orbital periods included in the given set; the fourth
column lists the value of $\beta_0$ used in the calculation, or the range in $\beta_0$
values; and the last column contains comments.  Unless otherwise noted, all calculations 
were performed with a kinematical viscosity parameter, 
$\nu$=0.005 R$^2_\odot$/day=2.8$\times$10$^{14}$ cm$^2$/s.

\begin{table}     
\caption{Summary of the 5$+$4 M$_\odot$ model binary system models for $\Delta R_1/R_1=$0.06 and
$\nu=$0.005 R$_\odot^2$/day (unless noted otherwise). \label{tbl-1}}
\label{table1}
\centering
\begin{tabular}{rlccl}
\hline\hline
Set Num. &$e$ & $P$     & ${\beta}_0$ &  \\
         &    &(days)   &             & Comments    \\
\hline
 1  & 0.0 & 2--265    &1.2, 2.0  & $\nu$=0.003 R$_\odot^2$ day$^{-1}$      \\
 2  & 0.0 & 2--265    &1.2       &                          \\
 3  & 0.1 & 3--255    &1.2       &                                  \\
 4  & 0.1 &   6       &0.4--2.1  & minimum $\dot{E}_{ave}$  at $\beta_0$=0.90  \\
 5  & 0.1 &   6       & 1.2      &   $\nu=$0.001--0.088 R$_\odot^2$ day$^{-1}$ \\
 6  & 0.1 &   6       & 1.2      &   $\Delta R_1/R_1$=0.02--0.10               \\
 7  & 0.3 & 3 --255   & 1.2      &                                    \\
 8  & 0.3 & 3 --390   & 2.0      &                                    \\
 9  & 0.3 &   6       & 1.2      &   $\Delta R_1/R_1$=0.02--0.10               \\
10  & 0.3 &   6       &0.4--2.35 &  minimum $\dot{E}_{ave}$  at $\beta_0$=0.80 \\
11  & 0.5 & 4--48     &1.2       &                                    \\     
12  & 0.5 &   6       &0.4--2.35 &minimum $\dot{E}_{ave}$ at $\beta_0$=1.05 \\ 
13  & 0.7 & 8--255    &1.2       &                                    \\  
14  & 0.8 & 13--350   &1.2       &                                    \\
15  & 0.8 & 20        &1.2       &  $\Delta R_1/R_1=$0.02--0.10               \\
16  & 0.8 & 20        &0.4--1.80 &minimum $\dot{E}_{ave}$ at $\beta_0\sim$1.15 \\   
17  & 0.8 & 30        &0.4--1.25 &minimum $\dot{E}_{ave}$ at $\beta_0\sim$1.15\\  
\hline
\end{tabular}
\end{table}

\section{Dependence of $\dot{E}_{ave}$ on layer depth}

\begin{figure}
\centering
\includegraphics[width=0.80\linewidth]{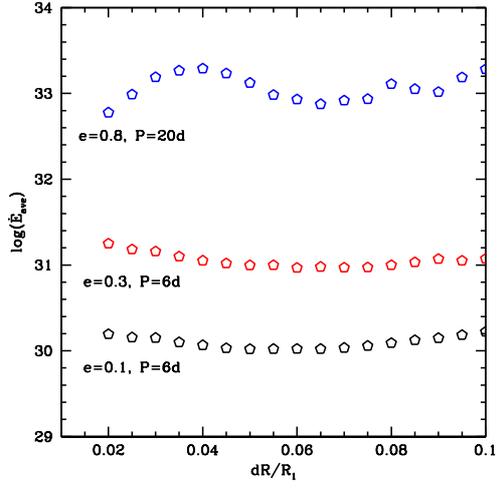}
\caption{Dependence of $\dot{E}_{ave}$ on the choice of layer depth  for  models
computed with e=0.1, 0.3 and e=0.8, and corresponding orbital periods as listed in
the label.  The average density for each layer depth used for the  $\dot{E}_{ave}$
calculation were obtained from the stellar structure model of a 5 Mo, 3.15 Ro star (I. Brott,
private communication, 2010). This shows that for a fixed set of binary parameters, the energy
dissipation rates depend on the choice of layer  depth only within a factor of $\sim$2.
}
\label{15874fg14}            
\end{figure}

We examined the manner in which the energy dissipation rates depend on the
choice of the depth ($\Delta R_1$/$R_1$) of the surface layer that is modeled.
This was done holding all parameters fixed, except for the values of d$R_1$/$R_1$,
which were varied from 0.02 to 0.10.  The results for $e$=0.1, 0.3 and 0.8 are
illustrated in Figure \ref{15874fg14}, which show that for the models we have run for this paper,
the calculations using different layer depths yield values of $\dot{E}_{ave}$ that 
differ by no more than a factor of two. 

\end{appendix}

\end{document}